\documentclass[aps, prd, amsmath, floats, floatfix, twocolumn,
superscriptaddress, nofootinbib, showpacs]{revtex4-1}
\usepackage[xetex]{graphicx}
\usepackage{color}
\usepackage{soul}
\usepackage{url}
\usepackage{bm}         % bold math symbols
\usepackage{times}

\newcommand{\beq}{\begin{equation}}
\newcommand{\eeq}{\end{equation}}
\newcommand{\beqn}{\begin{eqnarray}}
\newcommand{\eeqn}{\end{eqnarray}}

\newcommand{\lo}{\mathrel{\raise.3ex\hbox{$<$}\mkern-14mu
    \lower0.6ex\hbox{$\sim$}}}
\newcommand{\go}{\mathrel{\raise.3ex\hbox{$>$}\mkern-14mu
    \lower0.6ex\hbox{$\sim$}}}

\usepackage{color}

\newcommand{\Caltech}{\affiliation{TAPIR, Walter Burke Institute for Theoretical Physics, MC 350-17,
    California Institute of Technology, Pasadena, California 91125, USA}}
\newcommand{\Cornell}{\affiliation{Cornell Center for Astrophysics and Planetary Science, 
Cornell University, Ithaca, New York, 14853, USA}}
\newcommand{\WSU}{\affiliation{Department of Physics \& Astronomy,
	Washington State University, Pullman, Washington 99164, USA}}
\newcommand{\CITA}{\affiliation{Canadian Institute for Theoretical 
    Astrophysics, University of Toronto, Toronto, Ontario M5S 3H8, Canada}}

\newcommand{\CIFAR}{\affiliation{Canadian Institute for Advanced Research, 180 Dundas St.~West, Toronto, ON M5G 1Z8, Canada}} %
\newcommand{\LBL}{\affiliation{Lawrence Berkeley National Laboratory,
1 Cyclotron Rd, Berkeley, CA 94720, USA}}
\newcommand{\NCSU}{\affiliation{Department of Physics, North Carolina State University, Raleigh, North Carolina 27695, USA}}
\newcommand{\AEI}{\affiliation{Max-Planck-Institut fur Gravitationsphysik, Albert-Einstein-Institut, D-14476 Golm, Germany}}
\usepackage{graphicx}% Include figure files
\usepackage{dcolumn}% Align table columns on decimal point
\usepackage{bm}% bold math
\usepackage{epsf}

\begin{document}

\title{Low mass binary neutron star mergers : gravitational waves and neutrino emission}

\author{Francois Foucart}  \LBL
\author{Roland Haas} \AEI %
\author{Matthew D. Duez} \WSU %
\author{Evan O'Connor} \NCSU
\author{Christian D. Ott} \Caltech 
\author{Luke Roberts}\Caltech 
\author{Lawrence E. Kidder} \Cornell %
\author{Jonas Lippuner} \Caltech
\author{Harald P. Pfeiffer} \CITA\CIFAR
\author{Mark A. Scheel} \Caltech %

\begin{abstract}
Neutron star mergers are among the most promising sources of gravitational waves for advanced ground-based detectors.
These mergers are also expected to power bright electromagnetic signals, in the form of short gamma-ray bursts,
infrared/optical transients powered by r-process nucleosynthesis in neutron-rich material ejected by the merger, and radio emission
from the interaction of that ejecta with the interstellar medium. Simulations
of these mergers with fully general relativistic codes are critical to understand the merger and post-merger gravitational
wave signals and their neutrinos and electromagnetic counterparts. In this paper, we employ the Spectral Einstein Code (SpEC) 
to simulate the merger of low-mass neutron star binaries
(two $1.2M_\odot$ neutron stars) for a set of three nuclear-theory based, finite temperature equations of state. 
We show that the frequency peaks
of the post-merger gravitational wave signal are in good agreement with predictions obtained from recent simulations using
a simpler treatment of gravity. We find, however, that only the fundamental mode of the remnant is excited for long periods of time: 
emission at the secondary peaks is damped on a millisecond timescale in the simulated binaries. For such low-mass systems, the remnant is a 
massive neutron star which, depending on the equation of state, is either permanently stable or long-lived (i.e.~rapid uniform rotation is
sufficient to prevent its collapse). We observe strong excitations of $l=2$, $m=2$ modes, both in the massive neutron star and in
the form of hot, shocked tidal arms in the surrounding accretion torus. We estimate the neutrino emission of the remnant
using a neutrino leakage scheme and, in one case, compare these results with a gray two-moment neutrino transport 
scheme. We confirm the complex geometry of the neutrino emission, also observed in previous simulations with neutrino leakage, 
and show explicitly the presence of important
differences in the neutrino luminosity, disk composition, and outflow properties between the neutrino leakage and transport
schemes.
\end{abstract}


\maketitle

\section{Introduction}
\label{sec:intro}

Compact binary mergers, including binary neutron stars (BNS), binary black holes (BBH), and neutron star-black hole
(NS-BH) mergers are the primary targets of ground based gravitational wave detectors such as Advanced 
LIGO~\cite{aLIGO2}, Advanced VIRGO~\cite{aVirgo2}, and KAGRA~\cite{kagra}. 
In the presence of at least one neutron star, the merger may also be accompanied by bright
electromagnetic emission, whose detection can provide valuable complementary information on the properties of the 
source, help characterize the merger environment, and provide better localization than available from the gravitational wave signal 
alone~(see e.g.~\cite{metzger:11}
for a review). 
Two of the most promising such
counterparts are short gamma-ray bursts (SGRBs)~\cite{moch:93,LK:98,Janka1999,Fong2013} and radioactively powered transients originating from r-process
nucleosynthesis in the neutron-rich matter ejected by the 
merger~\cite{1976ApJ...210..549L,Li:1998bw,Roberts2011,Kasen:2013xka,Tanaka:2013ana}. The latter would most likely peak in 
the infrared about a week after merger~\cite{2013ApJ...775...18B,Tanaka:2013ana,Lippuner2015}. 
It could also result in the production of many of the 
heavy elements ($A\gtrsim 90$) whose origin remains poorly understood today~\cite{korobkin:12,Roberts2011,Wanajo2014}.
Finally, the ejecta could power long duration radio emission as it interacts with the interstellar medium.

Numerical simulations of these mergers play an important role in the efforts to model the gravitational wave signal,
predict the properties of its electromagnetic counterparts, and estimate the production of various elements.
In this work, we focus on the outcome of BNS mergers.
The simulation of BNS mergers with general relativistic hydrodynamics codes
has now been possible for about 15 years~\cite{Shibata00b}. Nevertheless, and despite continuous improvements, 
current codes have not yet reached the accuracy required
to model the gravitational wave signal at the level required to extract as much information as possible from upcoming
experiments (see e.g.~\cite{Barkett2015}).
Additionally, most codes do not take into account all of the physics relevant to the evolution of
the post-merger remnant, including at least a hot nuclear-theory based equation of state, 
a neutrino transport scheme accounting for neutrino-matter and neutrino-neutrino interactions~\cite{Sekiguchi:2015,Foucart:2015a}, 
and the evolution of the magnetic fields with a high enough resolution to resolve the growth of 
magnetohydrodynamics (MHD) instabilities~\cite{Kiuchi2014,Kiuchi2015}. 

Many recent simulations have taken steps toward an improved treatment of all relevant physics,
and a better coverage of the available parameter space of BNS configurations. 
Fully general relativistic simulations of the merger of two $1.35M_\odot$ neutron stars 
with an approximate neutrino transport scheme, but no magnetic fields~\cite{Sekiguchi:2015}, have shown that 
neutrino-matter
interactions can lead to the ejection of material with a broad distribution of temperature and composition, leading to
the production of elements with abundances compatible with solar system observations~\cite{Wanajo2014}.
The study of a similar merger at an unprecedented resolution with ideal MHD but using a simpler
equation of state and no neutrinos~\cite{Kiuchi2014} demonstrated multiple mechanisms for the growth of magnetic field
in global simulations of mergers -- although, even at the highest resolution currently achieved, the amplification of the magnetic
field in the shear region between the two merging neutron stars is not resolved~\cite{Kiuchi2015}, 
and the unresolved growth of magnetic field in that region could be very significant~\cite{2014arXiv1410.0013G,Kiuchi2015}.
Lower resolution simulations including {\it resistive} MHD have shown that deviations
from ideal MHD have a relatively modest effect on the evolution of the post-merger remnant~\cite{2015arXiv150202021D}. 
With or without resistivity, Dionysopoulou et al.~\cite{2015arXiv150202021D} also find low-density,
magnetically dominated outflow regions which may lead to the formation of a relativistic jet. 
The combined effect of ideal MHD with a subgrid model for the growth of the magnetic field, and a simple neutrino 
cooling scheme has also been investigated~\cite{Neilsen:2014hha,Palenzuela2015}. These results suggest that the growth
of strong magnetic field could influence the mass and properties of the outflows, and the dynamics of the post-merger
remnant. 

Large parameter space studies without 
microphysics~\cite{hotokezaka:13,Hotokezaka2013,Bernuzzi:2014kca,Takami:2014zpa,Takami:2015,DePietri2015} have also greatly 
improved our understanding of the general properties of the post-merger remnant. 
If the post-merger remnant collapses to a black hole, the relevant timescale for that collapse can be estimated from the total mass of the
binary. Order of magnitude estimates are available for the mass ejection from the merger, and we are beginning to understand the main features of the
post-merger gravitational wave signal.  More exotic merger scenarios, leading to more extreme mass ejection, merger results,
and orbital evolution have also been investigated, including eccentric binaries~\cite{Gold:2011df,East:2012ww}, and binaries with rapidly rotating neutron stars~\cite{Bernuzzi:2013rza,Kastaun:2013mv,Kastaun:2014fna,Dietrich:2015pxa,Tacik2015}.
Finally, a large number of simulations have also been performed using approximate treatments of gravity, producing
useful predictions for the properties of the post-merger gravitational wave signal~\cite{Bauswein:2014qla,Bauswein:2015a},
as well as the long term evolution of the post-merger 
remnant~\cite{Fernandez2013,2013ApJ...773...78B,Just2014,Perego2014,Fernandez:2014}. 
All of these, and many more earlier results, provide us with a growing understanding of the behavior of neutron star 
mergers (see also~\cite{Duez:2009yz,ShibataTaniguchi:2011,Lehner2014} for reviews of earlier results), but certainly do not 
yet draw a complete picture.

In this paper, we focus on an often neglected portion of the parameter space in general relativistic simulations:
neutron star binaries at the very low end of the expected mass distribution~\cite{Ozel2012}. We consider two $1.2M_\odot$ neutron
stars, and evolve them using the SpEC code~\cite{SpECwebsite}, a set of three different hot, nuclear theory based equations of state, 
and either a neutrino cooling scheme or a gray two-moment neutrino transport scheme capable of properly taking into
account neutrino-matter interactions~\cite{Foucart:2015a}. 
A similar system has been evolved using ideal MHD and a simpler equation
of state (leading, in particular, to an unrealistically large neutron star radius) in~\cite{2014arXiv1410.0013G}, with the aim to study the
growth of magnetic field, and with similar equations of state but approximate gravity and no neutrinos in~\cite{bauswein:12}. 
Nuclear theory-based equations of state and a neutrino cooling scheme have also recently been used for higher mass 
systems in~\cite{Neilsen:2014hha,Palenzuela2015}, providing us with a useful point of comparison for many of the 
qualitative features observed in our simulations. The post-merger gravitational wave signal can also be compared to numerical results
using nuclear theory-based equations of state but an approximate treatment of gravity~\cite{Bauswein:2014qla,Bauswein:2015a}.

We describe our numerical methods in Sec.~\ref{sec:methods}, and initial conditions in Sec.~\ref{sec:ID}.
Our simulations allow us to address a number of important
questions. First, in Sec.~\ref{sec:gw},
 we study the post-merger gravitational wave signal of low-mass BNS systems with hot nuclear-theory based equations of state. 
This allows us to compare our results with predictions coming
from simulations using a more approximate treatment of gravity. Those simulations found clearly marked peaks in the gravitational
wave signal at frequencies easily connected with the neutron star equation of 
state~\cite{Bauswein:2014qla,Bauswein:2015a}. We can also compare our results to differing predictions
coming from general relativistic simulations using simpler equations of state and more massive neutron 
stars~\cite{Takami:2014zpa,Takami:2015}.
The post-merger signal is expected to depend on the equation of state of the neutron star, and its qualitative properties vary 
with the total mass of the system. 
Thus, our results using general
relativistic simulations, a nuclear-theory based equations of state, and low-mass neutron stars usefully complement
existing studies in general relativity~\cite{Hotokezaka2013,Takami:2014zpa,Takami:2015} and approximate 
gravity~\cite{Bauswein:2015a}.

Second, in Sec.~\ref{sec:remnant}, we study the qualitative features of the remnant:
its density, temperature, composition, rotation profile, and the excitation of $l=2$, $m=2$ mode in the post-merger neutron star
and the accretion torus. We compare our results with simulations of higher mass systems using a similar level of physical 
detail~\cite{Palenzuela2015}. 

Third,
in Sec.~\ref{sec:neutrinos}, we analyze neutrino emission, and for the simulation in which a 
transport scheme is used, directly
assess the impact of using the simpler neutrino cooling method, determine the changes in composition due to
neutrino-matter interactions, and estimate how these changes affect the properties of the outflows 
(Sec.~\ref{sec:outflows}). This is the first time that such a comparison is possible in binary neutron star mergers
(we already performed a similar study on the accretion disk formed in a NS-BH merger~\cite{Foucart:2015a}).
This provides us with better insights into the limits of the simple cooling scheme. We can also qualitatively 
compare our results
with the small set of existing neutron star merger simulations using a similar neutrino transport 
scheme~\cite{Sekiguchi:2015}.

\section{Numerical Methods}
\label{sec:methods}

\subsection{Evolution equations}

We evolve Einstein's equations and the general relativistic equations of ideal hydrodynamics 
using the Spectral Einstein Code (SpEC)~\cite{SpECwebsite}. 
SpEC evolves those equations on two separate grids: a pseudospectral
grid for Einstein's equations, written in the generalized harmonic formulation~\cite{Lindblom:2007}, and a finite
volume grid for the general relativistic equations of hydrodynamics, written in conservative form.
The latter makes use of an approximate Riemann solver (HLL~\cite{HLL}) and high-order shock
capturing methods (fifth order WENO scheme~\cite{Liu1994200,Jiang1996202}), resulting in a second-order accurate
evolution scheme. For the time evolution, we use a third-order Runge-Kutta algorithm.
Finally, after each time step, the two grids communicate the required source terms, using a third-order
accurate spatial interpolation scheme. Those source terms are the metric and its derivatives (from the pseudospectral grid
to the finite volume grid) and the stress-energy tensor of the fluid (from the finite volume grid to the 
pseudospectral grid). In our current scheme, the radiation stress-energy tensor does not self-consistently
feed back onto the evolution of Einstein's equations. Direct measurements of the energy in the neutrino sector shows that the radiation energy density
is everywhere negligible as far as gravitational interactions are concerned. The neutrino stress-energy tensor is, however, fully coupled to the general
relativistic equations of hydrodynamics.
More details about these numerical methods can be found in~\cite{Duez:2008rb,Foucart:2013a}.

While the GR-hydrodynamics in SpEC was largely developed for the evolution of NS-BH mergers,
most of the algorithm carries over to the evolution of binary neutron star mergers. The main differences
are in the choice of the evolution gauge, and in the grid structure (Sec.~\ref{sec:grid}). 
In the generalized harmonic formalism,
the evolution of the coordinates follows the wave equation
\beq
g_{ab} \nabla^c\nabla_c x^b = H_a({\bf x})\,,
\eeq
where $g_{ab}$ is the spacetime metric, and $H_a({\bf x})$ an arbitrary function. Before merger, we make
the choice
\beq
H_a (x^i_c,t) = H_a(x^i_c,0) \exp{\left(-\frac{t^2}{\tau^2}\right)}\,,
\eeq
with $\tau = \sqrt{d_0^3/M}$, $d_0$ the initial separation, and $M$ the total mass of the binary at infinite separation.
Here, $x^i_c$ are the comoving spatial coordinates (which follow the rotation and inspiral of the binary)
and $H_a(x^i_c,0)$ are the value of $H_a$ at the initial time, chosen so that 
$\partial_t \alpha(x^i_c,t)=\partial_t \beta^i(x^i_c,t) = 0$. The lapse $\alpha$ and shift $\beta^i$ are
obtained from the standard 3+1 decomposition of the metric,
\beq
g_{ab} = -\alpha^2 dt^2 + \gamma_{ij} (dx^i + \beta^i dt)(dx^j + \beta^j dt)\,,
\eeq
where $\gamma_{ij}$ is the 3-metric on a slice of constant $t$. After a transient on the timescale $\tau$,
the gauge settles into the "harmonic gauge" $H_a=0$ (see~\cite{Lindblom2006} for a discussion of the uses of that gauge). 
During merger, we instead transition to the
"damped harmonic" gauge condition~\cite{Szilagyi:2009qz},
\beq
H^{\rm DH}_a = \left[\log{\left(\frac{\sqrt{\gamma}}{\alpha}\right)}\right]^2
\left[\frac{\sqrt{\gamma}}{\alpha}t_a - \gamma_{ai}\frac{\beta^i}{\alpha} \right]\,,
\eeq
where $\gamma$ is the determinant of the 3-metric $\gamma_{ij}$. We do this according
to
\beq
H_a(t) = H^{\rm DH}_a \left[ 1-\exp{\left(\frac{-(t-t_{DH})^2}{\tau_m^2}\right)}\right]\,,
\eeq
with $t_{\rm DH}$ the time at which we turn on the damped harmonic gauge, discussed in Sec.~\ref{sec:grid},
and $\tau_m = 100M$.

We choose $H_a$ based on what we found most efficient in NS-BH mergers,
in terms of the resolution required to reach a given accuracy. There, we find that the harmonic
gauge is a better choice for the evolution of neutron stars during inspiral.
The damped harmonic gauge, on the other hand, is favored for the evolution of black holes
and for very dynamical spacetimes. For the collapse of a merger remnant into a black hole, different
gauge choices are necessary (R. Haas et al., in prep.). However, we do not have to take this into consideration here, 
since the low-mass systems that we study do not collapse to a black hole within the duration of the simulation.

\subsection{Grid setup}
\label{sec:grid}

\begin{figure}
\flushleft
\includegraphics[width=1.\columnwidth]{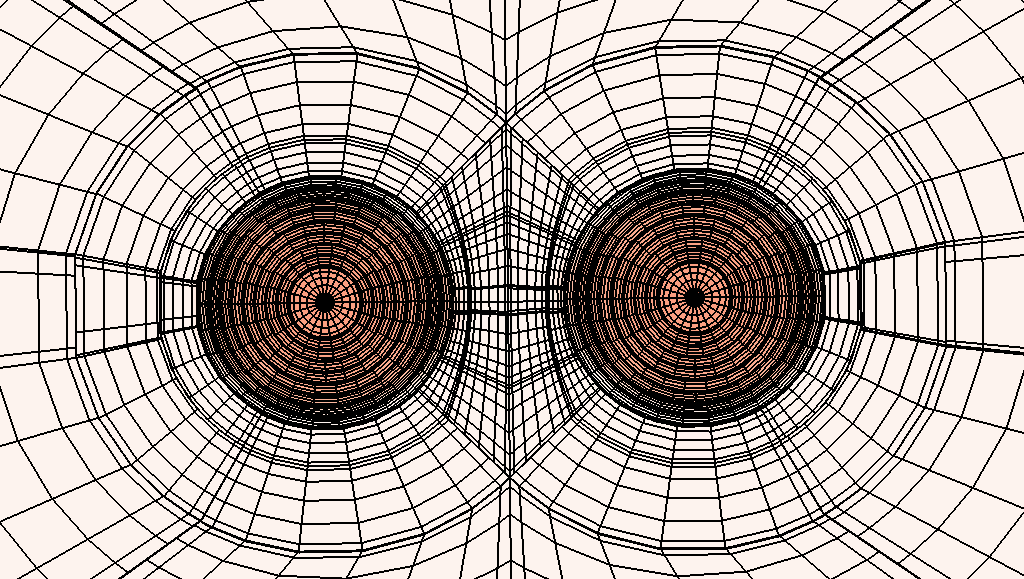}
\caption{Equatorial cut of the pseudospectral grid just before the two neutron stars get into contact. The color scale
shows the baryon density (logarithmic scale).}
\label{fig:InspiralGRGrid}
\end{figure}

Before the two neutron stars enter into contact, the pseudospectral grid on which we evolve Einstein's equations
takes advantage of the approximate
spherical symmetry in the neighborhood of each star, and in the far-field region. The evolved spatial slice
is decomposed into two small balls around the center of each neutron star, sets of spherical shells around
each star and in the far-field region, and distorted cubes to connect the three spherically symmetric regions. 
The grid decomposition used in our simulations is pictured in Fig.~\ref{fig:InspiralGRGrid}.
The inner ball is expanded into Zernike polynomials~\cite{Matsushima-Marcus:1995,1997JCoPh.136..100V}, the shells into Chebyshev polynomials (in radius) and spherical harmonics (in angle), 
and the distorted cubes in Chebyshev polynomials. The grid follows the centers of the
neutron stars, defined as the center of mass of the matter in the $x<0$ and $x>0$ half planes, 
through a simple rotation and scaling of the grid coordinates.\footnote{Here and in the rest of the paper, the center of mass is
defined in the coordinates of the simulation, and is clearly a gauge-dependent quantity. It is, however, a convenient quantity to
use to define the map between the grid and physical coordinates and keep the neutron stars approximately fixed on the grid.}

\begin{figure}
\flushleft
\includegraphics[width=1.\columnwidth]{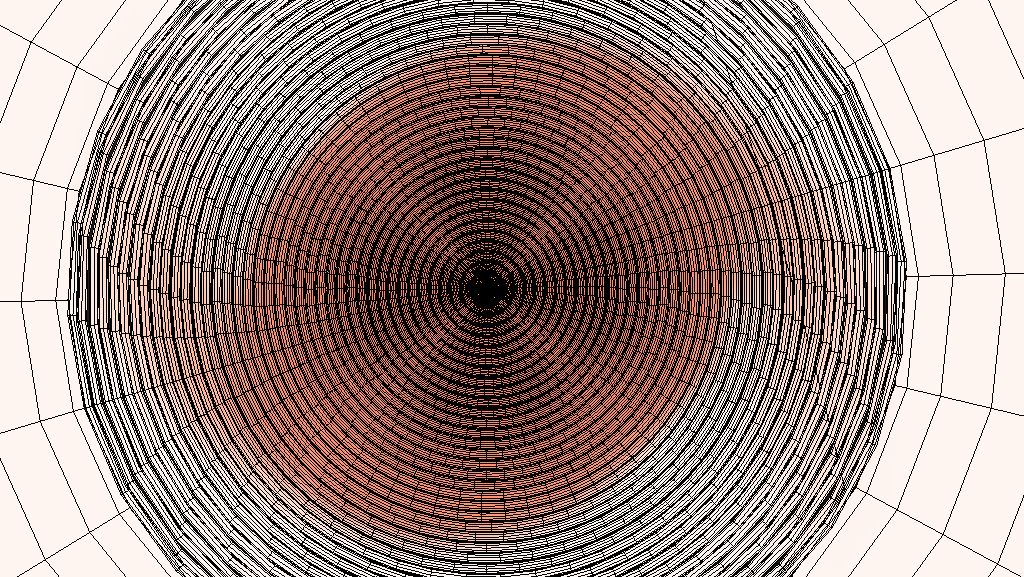}
\caption{Equatorial cut of the pseudospectral grid just after the change to a grid centered on the post-merger
neutron star remnant. The color scale shows the baryon density (logarithmic scale).}
\label{fig:MergerGRGrid}
\end{figure}

We maintain this grid decomposition for the evolution of Einstein's equations up to the point at which the maximum density on the grid increases
beyond the low-level oscillations observed during the inspiral. This rise in the density signifies the transition from
two well-separated neutron star cores to a single, more massive object. At that point, we switch to a grid
which is fully centered on the coordinate center of mass of the system. This grid, pictured in Fig.~\ref{fig:MergerGRGrid},
is made of a ball at the origin of the coordinate system, surrounded by 59 spherical shells extending to the outer
edge of the computational domain. Both before and after merger, the outer boundary is located at $40d_0$,
with $d_0$ the initial separation of the binary. For the configurations considered here, the outer boundary is 
thus at a coordinate radius of $400M-435M \sim (1400-1530)\,{\rm km}$.

Both before and after merger, the resolution on the pseudospectral grid is chosen adaptively, in order to reach a
target accuracy estimated from the convergence of the spectral expansion of the 
solution~\cite{Szilagyi:2014fna,Foucart:2013a}. As in
NS-BH mergers, our standard choice is to request a relative error of $10^{-4}$ in most of the
computational domain, with a smooth transition to $10^{-5}$ close to the center of each of the neutron stars
before the neutron stars get into contact (see~\cite{Szilagyi:2014fna,Foucart:2013a} for more detail on how this
is computed in practice). Larger and lower truncation errors can be imposed for convergence tests,
chosen in order to converge faster than the expected second-order convergence of the finite volume code. 

The finite volume grid on which we evolve the general relativistic equations of hydrodynamics is very simple.
Before the two neutron stars get into contact, it is composed of two cubes, each
with $96^2\times 48$ cells each and centered on one neutron star. The lower number of cells in the vertical dimension
is due to the assumption of equatorial symmetry before merger. The initial extent of the boxes is listed for each simulation in 
Table~\ref{tab:grid} -- it is chosen to cover a little more than the original size of the neutron star. In the coordinate system
comoving with the neutron star centers, the neutron stars expand as the binary inspirals. To avoid losing matter to the outer
boundary of the finite volume grid, we expand the grid by $4.5\%$ every time the flux of matter across the outer boundary
exceeds $0.015\,M_\odot s^{-1}$. As the inspiral lasts less than $10\,{\rm ms}$, this implies a mass loss well below 
$10^{-4}M_\odot$ before merger. 

As the two neutron stars approach each other, the two finite volume boxes will eventually get into contact. During merger, we
would like to follow the forming massive neutron star remnant, the tidal tails, the accretion disk, and any ejected material.
We switch to a finite volume grid centered on the forming remnant and with $2-3$ levels of refinement. Each level
has, at our standard resolution, $200^2\times 100$ cells, with the finest grid spacing listed in Table~\ref{tab:grid} and each 
coarser level increasing the grid spacing by a factor of 2.
The lower number of cells in the vertical direction now reflects
the fact that the remnant is less extended in that direction, and thus that we do not need the finest grid to extend as far vertically
as horizontally. During merger, we no longer assume equatorial symmetry. This numerical resolution, although 
insufficient to capture the evolution of magnetic fields~\cite{Kiuchi2014,Kiuchi2015}, 
has been shown to be sufficient to obtain reasonable accuracy in purely hydrodynamic simulations~\cite{Palenzuela2015}. 
We estimate the errors by performing a simulation with $256^2\times 128$ cells at each 
refinement level during merger ($122^2\times 61$ during the inspiral). These error estimates will be discussed along the relevant results in the remainder of this paper.

For these first simulations of neutron star mergers with microphysics and a neutrino leakage scheme in SpEC, 
we only use two levels of refinement 
during merger. This is far from ideal, as it allows us to follow
material only up to slightly more than $100\,{\rm km}$ from the central remnant ($50\,{\rm km}$ in the vertical direction). The
reason for this choice is simply to keep the simulations computationally affordable. With two levels of refinement, the mergers use
about $500$ cores for $16-20$ days, or about $200,000$ CPU-hours. Although not particularly large compared to some
of the most advanced general relativistic magnetohydrodynamics simulations performed to 
date~\cite{Kiuchi2014,Palenzuela2015,Kiuchi2015}, this reaches the limits of
our computational resources. An important part of this work is to demonstrate the ability of the SpEC code
to perform BNS mergers with microphysics, and to run efficiently on a larger number of cores than used
in previous studies (we typically use $\sim 50-100$ cores for NS-BH mergers). With our relatively small finite volume grid,
we can follow the formation of the post-merger neutron star remnant. We can also extract the characteristics of the remnant
and of the surrounding accretion disk, observe the gravitational wave signal\footnote{The pseudospectral grid on which we evolve
Einstein's equation extends to much larger distances.}, and study neutrino effects. The mass of the 
outflows, on the other hand, is very approximate. 

For the simulation using neutrino transport, we are however more interested in the properties of the outflows. 
And the composition of the ejected matter varies as it moves away from the disk, due to neutrino absorption. Accordingly, for that simulation, we choose to use a third level of refinement (i.e.~we make the grid twice as large). 

\subsection{Neutrino treatment}

Most simulations presented here are performed using a leakage scheme to capture the cooling and composition
evolution of the post-merger remnant due to neutrino emission. Our leakage scheme is described in more detail
in~\cite{Deaton2013,Foucart:2014nda}, and is based on previous work in Newtonian 
theory~\cite{Rosswog:2003rv,1997A&A...319..122R} and core-collapse supernovae~\cite{OConnor2010}. 
Effectively, it is a prescription for the energy and number of neutrinos leaving a given point of the remnant
as a function of the local properties of the fluid and the optical depth between that point and the outer boundary of the
computational domain (computed as in~\cite{Neilsen:2014hha}). 
The leakage scheme computes two rates of emission: a free-emission rate,
which only depends on the local properties of the fluid and is valid in the optically thin regime, and an approximate, steady
state  diffusion rate which depends on the optical depth and is accurate within an order of magnitude in the optically thick regime. 
In regions of moderate 
optical depth, an interpolation
formula between those two emission rates is used by the leakage scheme. 
For the emission and absorption rates, we use the values derived by 
Rosswog \& Liebendorfer~\cite{Rosswog:2003rv} for the charged-current reactions (which only affect electron neutrinos 
and antineutrinos),
\beqn
p + e^- &\leftrightarrow& n + \nu_e,\\
n + e^+ &\leftrightarrow& p + \bar\nu_e,
\eeqn
electron/positron pair creation and annihilation,
\beqn
e^+ + e^- &\leftrightarrow& \nu + \bar\nu,
\eeqn
and plasmon decay,
\beqn
\gamma &\leftrightarrow& \nu + \bar\nu.
\eeqn
We also use the rates of Burrows et al.~\cite{Burrows2006b} for nucleon-nucleon Bremsstrahlung
\beq
N + N \leftrightarrow N + N + \nu + \bar\nu.
\eeq
Finally, we use the rates of Rosswog \& Liebendorfer~\cite{Rosswog:2003rv} for the scattering of neutrinos on protons, 
neutrons and heavy nuclei.

To determine the errors in the leakage scheme and obtain more accurate information about neutrino effects,
we also perform a single simulation using a more advanced neutrino transport method. We use the
moment formalism, in which the energy-integrated (gray) energy density and momentum density of neutrinos is evolved on the 
grid~\cite{1981MNRAS.194..439T,2011PThPh.125.1255S,Cardall2013}.
An analytical closure (M1) is then used to compute the neutrino pressure tensor and close the system of equations.
The details of our M1 algorithm can be found in~\cite{Foucart:2015a}. Beyond providing more accurate results and information about the
spatial distribution of neutrinos, the transport scheme also allows us to take into account the absorption of neutrinos
and antineutrinos in low-density regions, which strongly affects the composition of the fluid, and particularly of the disk
and ejecta. These effects were shown to be significant in black hole-neutron star mergers~\cite{Foucart:2015a}. 
By explicitly comparing
leakage and transport results, we will show here that using a transport scheme is also critical for extracting the composition 
of the ejecta of binary neutron star mergers. So far, only a few simulations have used a similar transport scheme for neutron
star mergers~\cite{Wanajo2014,Sekiguchi:2015}. The transport scheme used in~\cite{Wanajo2014,Sekiguchi:2015} could, however, 
differ significantly from our algorithm in optically thick
regions. Indeed, it relies partially on rates from the leakage scheme to compute the evolution of the component of the neutrino
stress-energy tensor which is not trapped by the fluid. We instead evolve both the trapped and non-trapped components
of the neutrino stress-energy tensor together within the moment formalism (but rely on the leakage scheme to estimate the
average energy of neutrinos in optically thin regions~\cite{Foucart:2015a}).

For both neutrino schemes, we consider three separate neutrino species: the electron neutrinos $\nu_e$, electron antineutrinos $\bar\nu_e$,
and a third species regrouping all heavy-lepton neutrinos and antineutrinos ($\nu_\mu$, $\bar\nu_\mu$, $\nu_\tau$, $\bar\nu_\tau$),
which we denote $\nu_x$. Grouping all heavy-lepton neutrinos and antineutrinos in one species is justified in our simulations because
the temperatures reached in BNS mergers are low enough to suppress the formation of the corresponding heavy leptons. Thus, we do not
have to take into account the charged-current reactions with muon and tau leptons which would require us to differentiate between
the four species regrouped in $\nu_x$.

Finally, we note that the computation of the average energy of the neutrinos differ from what was presented in our previous
simulations~\cite{Deaton2013,Foucart:2014nda,Foucart:2015a}. This is due to the discovery of an error in the computation of the
average energy of neutrinos coming from optically thick regions in our leakage code, and in the GR1D code upon which it is based.
This error leads to an overestimate of the neutrino energies. 
Interestingly, because the post-merger remnants in black hole-neutron star and
binary neutron star systems are in a quasi-equilibrium state in which the neutrino luminosities and electron fraction of the fluid
are adapting to the evolving condition in the remnant (density, temperature) on a timescale shorter than the dynamical timescale
of the remnant, we find that correcting this error does not significantly affect the composition evolution, the neutrino luminosities, or
the dynamics of the remnant. It does not even noticeably affect the absorption rate of neutrinos in optically thin regions when 
coupling our neutrino leakage and transport schemes, because the computation of the neutrino temperature used to determine
the absorption rate in the transport scheme was previously corrected in optically thick regions to match known test results in
spherical symmetry (as described in~\cite{Foucart:2015a}). This error in the neutrino energies was discovered after the 
completion of the simulations presented here. We have performed significant portions of the simulations 
anew ($\sim 3\,{\rm ms}$), to verify that, as stated, our results are not significantly modified by the use of an improved estimate 
of the neutrino energies. The neutrino energies quoted in this work use the improved estimate of the neutrino spectrum in 
optically thick regions, and are typically $\sim 50\%$ lower than before corrections. Updated estimates of the neutrino energies in 
the systems considered  in previous publications using the SpEC code~\cite{Deaton2013,Foucart:2014nda,Foucart:2015a} 
will be provided separately.

\section{Initial Conditions and Equations of State}
\label{sec:ID} 

We obtain quasi-equilibrium initial conditions for neutron star binaries by solving the constraints in Einstein's equations
using the spectral elliptic solver Spells~\cite{Pfeiffer2003}. 
The algorithm used to generate initial data for BNS is similar 
to that previously developed for NS-BH binaries~\cite{FoucartEtAl:2008}, and spinning neutron star
binaries~\cite{Tacik2015}, and will be discussed in more detail in 
an upcoming paper (Haas et al., in prep). We consider three different physical configurations. Their main properties are listed in 
Table~\ref{tab:grid}. In each case, the two neutron stars have equal gravitational masses in isolation, $M_{\rm NS}=1.2M_\odot$,
chosen to probe the low-end of the expected mass distribution function of neutron stars in compact binaries~\cite{Ozel2012}. 
The neutron
stars are initially non-spinning, and on low eccentricity orbits: our initial data solver generates binaries with 
eccentricities of a few percents. Lower eccentricities can be achieved through an iterative procedure requiring the evolution
of the system for 2-3 orbits in each iteration~\cite{Pfeiffer-Brown-etal:2007}. 
However, this procedure would not perform well at the small initial separations considered here.

\begin{figure}
\flushleft
\includegraphics[width=1.03\columnwidth]{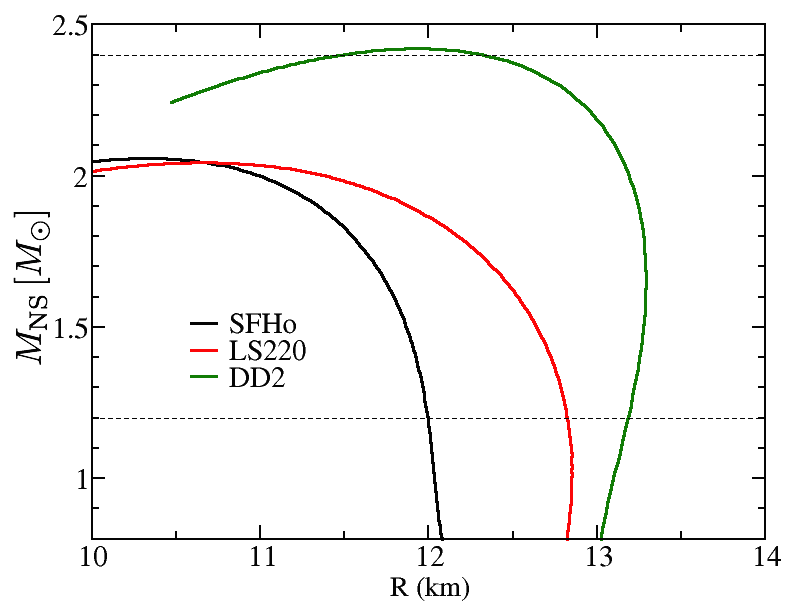}
\caption{Mass radius relationships for the three equations of state considered here, for cold neutron stars in neutrino-less $\beta$-equilibrium.
 Horizontal dashed lines are provided 
at the initial mass of the individual neutron stars ($1.2M_\odot$) and total mass of the binaries ($2.4M_\odot$)
considered here.}
\label{fig:MR}
\end{figure}

We consider three different, tabulated, temperature dependent nuclear-theory 
based equations of state, obtained from www.stellarcollapse.org 
(see~\cite{OConnor2010} for a description of the tabulated equations of state). 
These equations of state are SFHo~\cite{Steiner:2012rk}, DD2~\cite{Hempel:2011mk}, and LS220~\cite{Lattimer:1991nc}.
All three were initially constructed with the objective to satisfy the known theoretical and experimental constraints from nuclear physics,
although the LS220 equation of state is now incompatible with some recent results~\cite{Fischer2014}.
All three equations of state can support neutron stars with $M_{\rm NS}\gtrsim 2M_\odot$, as required
by astronomical observations~\cite{Demorest:2010bx,2013Sci...340..448A}. 
The mass-radius relationships for these equations of state are provided on Fig.~\ref{fig:MR} for cold, neutrino-less
$\beta$-equilibrium neutron
stars. The radii seen on Fig.~\ref{fig:MR} for $1.2M_\odot$ neutron stars are within
the range deemed most likely by studies incorporating information about both nuclear physics and astrophysical observations
of neutron stars in quiescent X-ray binaries~\cite{Steiner2010,2013ApJ...765L...5S}. 
A different analysis of the same astrophysical data however predicts
significantly more compact neutron stars~\cite{Guillot:2013wu}, and the physical compactness of neutron stars remains an 
important open question today.
Before merger, the binaries considered here go through 2.5 (LS220), 3 (DD2) and 4.5 (SFHo) orbits.

\begin{table}
\caption{
  Initial conditions and grid settings for the binary neutron star mergers studied here.
  ``EoS'' is the equation of state of the neutron star matter, $R_{\rm NS}^{1.2M_\odot}$ the
  radius of a neutron star of gravitational mass $M_{\rm NS}=1.2M_\odot$, $M_b^{1.2M_\odot}$
  the baryonic mass of the same neutron star, $\kappa_2^{1.2M_\odot}$ is the tidal
  coupling constant (see text and e.g.~\cite{Bernuzzi2015})\footnote{The authors thank Jan Steinhoff for providing the values 
  of $\kappa_2$ for these equations of state.}, $d_0$ the initial coordinate separation,
  $\Delta x_{\rm FD}^{t=0}$ the initial coordinate grid spacing on the finite volume grid, and
  $\Delta x_{\rm FD}^{\rm merger}$ the grid spacing in the finite volume grid used after the formation
  of a supermassive neutron star.
}
\label{tab:grid}
\begin{ruledtabular}
\begin{tabular}{|c|c|c|c|c|c|c|}
{\rm EoS} & $R_{\rm NS}^{1.2M_\odot}$ & $M_b^{1.2M_\odot}$ & $\kappa_2^{1.2M_\odot}$ & $d_0$ & $\Delta x_{\rm FD}^{t=0}$ & $\Delta x_{\rm FD}^{\rm merger}$\\
\hline
LS220 & 12.8 km & 1.309 & 271 & 35.2 km & 252 m & 306 m\\
 LS220 & 12.8 km & 1.309 & 271 & 35.2 km & 198 m & 234 m\\
DD2 & 13.2 km & 1.295 & 307 & 36.6 km & 277 m & 322 m\\
SFHo & 12.0 km & 1.303 & 164 & 38.3 km & 249 m & 276 m\\
\end{tabular}
\end{ruledtabular}
\end{table}

The stiffest equation of state, DD2, has a maximum mass above $2.4M_\odot$ even for cold, non-rotating
neutron stars. The merger of two $1.2M_\odot$ neutron stars will thus result in a stable remnant.
For the other two equations of state, the merger remnant will eventually collapse into a black hole. 
We will see that a massive neutron star remnant does, however, survive for at least $10\,{\rm ms}$ after the merger.
Note that the maximum baryon mass of a cold,
uniformly rotating neutron star (as determined from mass-shedding sequences
generated by the code of Cook, Shapiro, and Teukolsky~\cite{cook92,cook94a}) is $2.83M_\odot$ for the LS220 equation of state, $2.86M_\odot$ for the SFHo equation
of state, and $3.45M_\odot$ for the DD2 equation of state. 
The total baryon masses of the systems under consideration are $2.62M_\odot$ (LS220), $2.61M_\odot$ (SFHo), and $2.59M_\odot$ (DD2).
Hence, for the LS220 and SFHo equations of state, the relevant timescale for the remnant to collapse to a black hole is 
the pulsar spin-down timescale, which is much longer 
than the timescales relevant to the study of BNS mergers or of the immediate post-merger remnant evolution (see 
also~\cite{Kaplan2013} for a more detailed discussion of the fate of the post-merger remnant in BNS systems).

\section{Gravitational Waves}
\label{sec:gw}

We extract the gravitational wave signal from the simulations following the method presented by 
Boyle \& Mroue~\cite{Boyle-Mroue:2008}:
the gravitational waves are extracted on coordinate spheres at 24 radii equally spaced in $r^{-1}$ 
within the range $[140,1325]\,{\rm km}$. We then
extrapolate the signal to infinity by fitting the finite-radius data with a second-order polynomial in $r^{-1}$. For the extrapolation,
we use finite radius values at the same retarded time (as defined in~\cite{Boyle-Mroue:2008}), to account for the finite propagation 
speed of gravitational waves. The extrapolation error can be estimated by comparing fits of different polynomial order to the finite 
radius data.

\subsection{Inspiral and tidal effects}

\begin{figure}
\flushleft
\includegraphics[width=1.03\columnwidth]{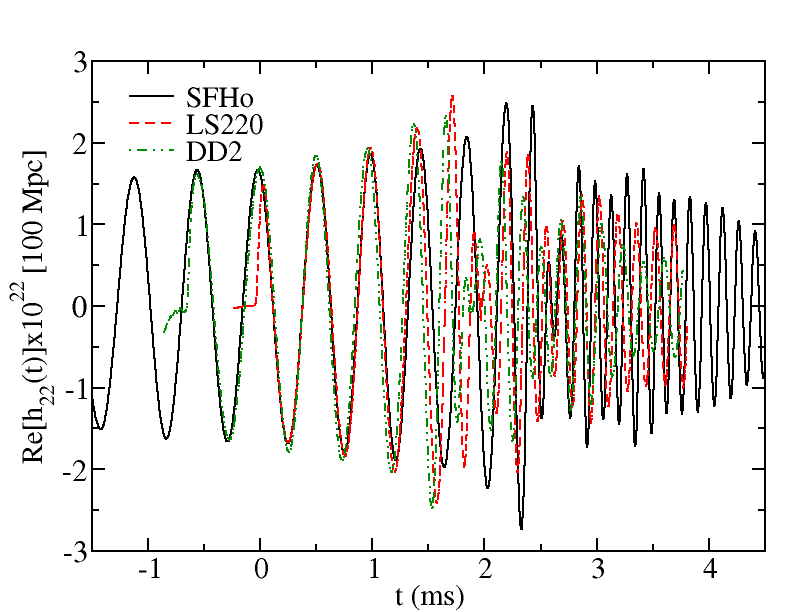}
\caption{Dominant (2,2) mode of the gravitational wave strain for all three configurations, at $100\,{\rm Mpc}$ from the source.
The simulation with the 
stiffest equation of state (DD2) shows the fastest inspiral. As opposed to other figures in this paper, we show the waveforms extrapolated
to infinity, which explains the short length of the post-merger waveform.}
\label{fig:waveforms}
\end{figure}

The purpose of this work is to study the merger and post-merger dynamics of neutron star binaries. 
The simulations performed here are
generally too short to perform detailed studies of the gravitational waveform during inspiral. We will focus more on
the post-merger signal in the next section. Qualitatively, however, we see that the late inspiral proceeds as expected.
Figure~\ref{fig:waveforms} shows the dominant mode of the gravitational wave signal, approximately matched in time and phase 
at the beginning of the LS220 simulation. 
We see that neutron stars with larger radii (LS220, DD2) inspiral faster than neutron stars with small radii (SDHo), and merge
at a lower frequency. The first effect is due to stronger tidal dissipation for neutron stars of larger radii. Tides cause a phase difference 
in the gravitational wave signal proportional to the tidal coupling constant $\kappa_2$ (see Table~\ref{tab:gw}), which is strongly
correlated with the radius of the neutron star.
The tidal coupling constant is, for an equal mass binary,
$\kappa_2 = k_2/(8C_{\rm NS}^5)$ with $C_{\rm NS}=GM_{\rm NS}/(R_{\rm NS}c^2)$ the compactness of an isolated neutron star,
and $k_2$ the dimensionless Love number. The second effect is due to the fact that the merger frequency depends on 
the size of the individual neutron stars, with more compact neutron stars merging at higher frequencies.

\subsection{Post-Merger gravitational wave signal}

The post-merger signal provides more useful information. Previous studies have shown that the gravitational wave
spectrum of the post-merger remnant of a neutron star binary shows clear peaks at frequencies dependent on the 
equation of state of the neutron 
star~\cite{1994PhRvD..50.6247Z,2002PhRvD..65j3005O,Shibata2002,Baiotti:2008ra,Kiuchi:2009jt,stergioulas:11,bauswein:12,Hotokezaka2013,Bernuzzi:2013rza,Bauswein:2014qla,Takami:2014zpa,Takami:2015}. 
The strongest of those peaks ($f_{\rm peak}$) occurs at the frequency 
of the fundamental quadrupole mode of the remnant. 
Recently, Bauswein \& Stergioulas~\cite{Bauswein:2015a} have associated the two largest 
potential secondary peaks with rotating spiral structures in the remnant ($f_{\rm spiral}$), and a coupling between the
fundamental quadrupole mode and quasi-radial oscillations in the remnant ($f_{2-0}$). They also provide a
fitting formula for the location of those peaks, based on simulations using an approximate treatment of gravity.
These simulations assume that the metric is conformally flat, an assumption which can accommodate exactly non-spinning,
isolated black holes and neutron stars, but not spinning compact objects or binary systems.
In recent simulation of BNS mergers for $1.35M_\odot$ neutron stars, with a general relativistic code
and nuclear-theory based equations of state, Palenzuela et al.~\cite{Palenzuela2015} find deviations 
of $\sim 0.1-0.3\,{\rm kHz}$ 
from the fits for $f_{\rm peak}$ and $f_{\rm spiral}$ provided in~\cite{Bauswein:2015a}. 
A minor peak is sometimes also found close to the predicted location of
$f_{2-0}$, but other features of similar amplitudes are also found at other frequencies.

\begin{figure}
\flushleft
\includegraphics[width=1.0\columnwidth]{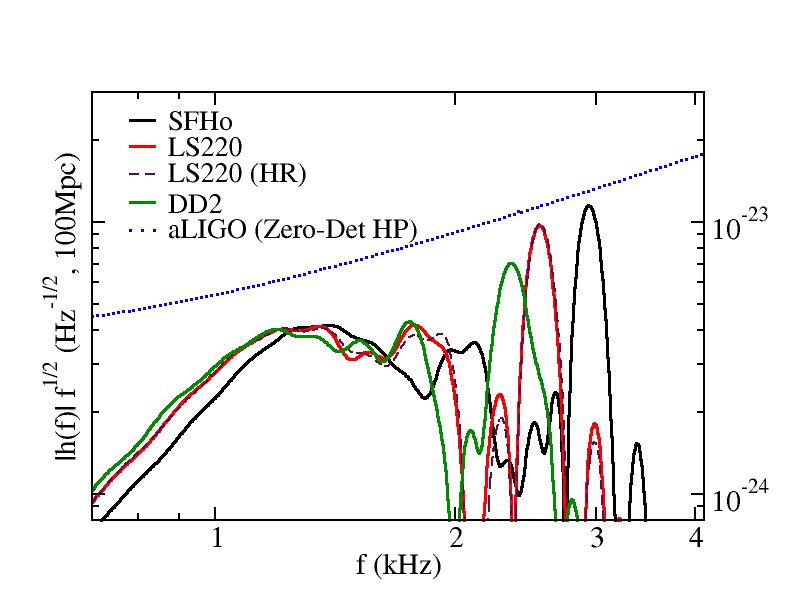}
\caption{Power spectrum of the merger and post-merger gravitational wave signal for optimally oriented mergers
at a distance of $100\,{\rm Mpc}$, multiplied by $f^{1/2}$.
For reference, we also plot the design sensitivity of advanced LIGO (Zero-Detuned High Power detector strain noise spectrum,
dashed blue curve)~\cite{Shoemaker2009}. Note the clear dominant peaks at $(2.3-3)\,{\rm kHz}$. All spectra
are Fourier transforms of the dominant (2,2) mode of the merger waveform in the time interval 
$[t_{\rm peak}-2\,{\rm ms},t_{\rm peak}+6\,{\rm ms}]$, with a tapering window of width $0.8\,{\rm ms}$ used at the beginning and
end of the interval.
}
\label{fig:spectra}
\end{figure}

The spectra of the $(2,2)$ mode of the gravitational wave signals observed in our simulations are shown in 
Fig.~\ref{fig:spectra}. For all equations
of state, emission at the fundamental mode is clearly visible. A number of secondary peaks are also observed. 
The post-merger signals are weak when compared, for example, to the design sensitivity of the Advanced LIGO
detector: a very close event is required in order to detect the post-merger signal~\cite{Clark2014}.
In Table~\ref{tab:gw},
the frequencies of the fundamental and strongest secondary peaks are compared with theoretical predictions. For the 
fundamental mode we use the fitting formula recommended by Bauswein et al. 2012~\cite{bauswein:12},
\beqn
f_{\rm peak}^{\rm fit} &=& (-0.2823 R_{1.6} + 6.284) \sqrt{\frac{2.4}{2.7}} \,\,[f_{\rm peak} < 2.64\,{\rm kHz}]\nonumber\\
&=& (-0.4667R_{1.6} + 8.713) \sqrt{\frac{2.4}{2.7}} \,\, [f_{\rm peak} > 2.64\,{\rm kHz}],\nonumber
\eeqn
with $R_{1.6}$ the radius of a neutron star of mass $M_{\rm NS}=1.6M_\odot$ [in kilometers], $f_{\rm peak}$
given in kHz, and we assume a total mass of $2.4M_\odot$.
Here, we used the proportionality relation $f_{\rm peak}\propto \sqrt{M_{\rm tot}/R_{\rm max}^3}$~\cite{bauswein:12},
with $M_{\rm tot}$ the total mass of the binary at infinite separation,
and $R_{\rm max}$ the radius of a neutron star at the maximum mass $M_{\rm max}$ allowed for this equation of state.
We find disagreements of only $\sim (30-70)\,{\rm Hz}$ between the numerical results and the fitting
formula, which would translate to systematic errors 
of $\sim (100-200)\,{\rm m}$ in the radius of a $1.6M_\odot$ neutron star.
We should also note that the merger of two $1.2M_\odot$ neutron stars with the LS220 equation of state was
studied with an approximate treatment of gravity in~\cite{bauswein:12}. 
The dominant frequency of the post-merger signal was $2.55\,{\rm kHz}$ in that study.
We find an extremely close value for that dominant frequency, $2.56\,{\rm kHz}$.

For the secondary peak, we compare our results to the linear fits to $f_{\rm spiral}$ and $f_{2-0}$ provided
in Fig. 4 of Bauswein \& Stergoulias~\cite{Bauswein:2015a}. Bauswein \& Stergoulias predict that the spiral mode 
should be the dominant secondary mode
for mergers in which a stable or long-lived hypermassive neutron star is formed. This is in good agreement with our results
as the first subdominant peak observed in our spectra agrees well with the frequency of $f_{\rm spiral}$.
The SFHo equation of state waveform also shows a peak close to $f_{2-0}$ and the LS220 equation of state waveform
has some extra power at $f_{2-0}$. For the DD2 equation of state waveform, the predicted location of the $f_{2-0}$ peak is too 
close to the merger frequency for any clear feature to be observed. We should note, however, that the predictions provided 
in~\cite{Bauswein:2015a} for $f_{\rm spiral}$ and $f_{2-0}$ are not as powerful as the unique fitting formula provided for
$f_{\rm peak}$. This is because a different linear relation between the frequency of the mode and the compactness of the star has
to be determined for each choice of neutron star masses. A universal relation between the secondary peak of the post-merger
waveform and the neutron star compactness has been proposed by Takami et al.~\cite{Takami:2014zpa,Takami:2015}.
This universal relation would predict that the secondary peak is at $\sim 1.6\,{\rm kHz}$ for the LS220 and DD2 equations of 
state and at $\sim 1.7\,{\rm kHz}$ for the SFHo equation of state. The SFHo has its third strongest peak at that frequency, and
that peak is not much weaker than the secondary peak. The other two equations of state show a difference of 
$(100-200)\,{\rm Hz}$ between the theoretical predictions and the location of the secondary peak, which would translate into 
$(0.5-1.0)\,{\rm km}$ errors in the determination of the neutron star radius. Using the universal formula
from~\cite{Takami:2014zpa,Takami:2015} to infer the compactness of the neutron stars considered in this paper would thus 
lead to significant errors in the determination of neutron star radii. Similar differences were observed by 
Bauswein \& Stergoulias~\cite{Bauswein:2015a} for low-mass binaries. Like 
Takami et al.~\cite{Takami:2014zpa,Takami:2015} (but as opposed to Bauswein \& Stergoulias~\cite{Bauswein:2015a}), 
our code is fully general relativistic. The observed differences are thus not due to the treatment of gravity. A more likely
explanation is the use of a lower mass system combined with the use of nuclear-theory based equations of state.
The choice of equation of state may also explain why Takami et al.~\cite{Takami:2014zpa} find significant differences
between the frequency of the fundamental mode and $f_{\rm peak}^{\rm fit}$, while we do not. The largest differences
in~\cite{Takami:2014zpa} were observed for the simple $\Gamma=2$ polytrope, while more realistic equations of state
performed better. The general relativistic simulations of Palenzuela et al.~\cite{Palenzuela2015}, performed for higher mass systems,
found $\sim 10\%$ disagreement between the numerical results and $f_{\rm peak}^{\rm fit}$ -- a larger difference than in our
simulations, but one that would still allow the recovery of the neutron star radius with systematic errors $\ll 0.5\,{\rm km}$.
It is quite likely that the fitted frequency $f_{\rm peak}^{\rm fit}$ is not universal, but nonetheless practically 
applicable to realistic neutron star equations of state.

\begin{figure*}
\includegraphics[width=1.\linewidth]{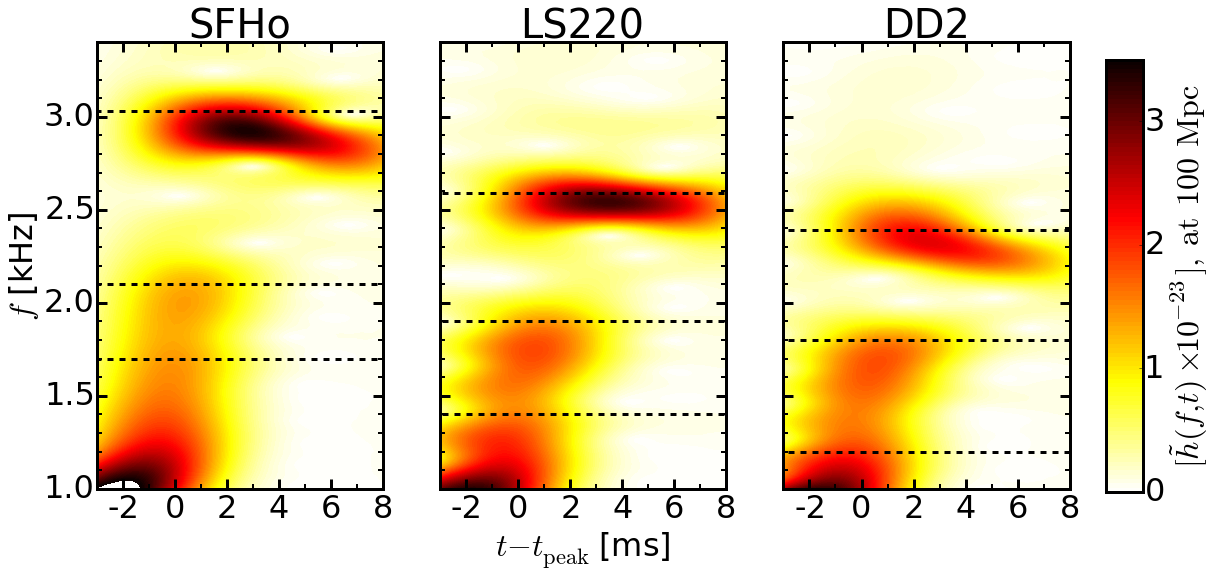}
\caption{Spectrograms of the gravitational wave signal for the simulations using the SFHo (left), LS220 (middle) and DD2 (right) equations of state, for an optimally oriented binary at $100\,{\rm Mpc}$. 
The dashed horizontal black curves show the location of the peaks predicted in~\cite{Bauswein:2015a}. The time is calculated as the difference between the center of the window function used
to compute the spectrogram (see text) and the peak of the gravitational wave amplitude. 
The dominant peak is clearly visible at $(2.3-3)\,{\rm kHz}$. Secondary peaks in the
$(1.3-2.1)\,{\rm kHz}$ range are poorly resolved, and quickly damped. The strong emission at frequencies
$f\lesssim 1.3\,{\rm kHz}$ is the merger signal itself.}
\label{fig:spectrogram}
\end{figure*}

More insight can be gained in those post-merger features by considering a spectrogram of the gravitational wave signal,
shown in Fig.~\ref{fig:spectrogram}. The quantity plotted there is
\beq
\tilde{h}(t_0,f) = \frac{|\int h_{2,2}(t) W(t-t_0) e^{ift} dt|}{\int W(t-t_0)dt},
\eeq
where we choose for the window function $W$ the exact Blackman window,
\beqn
W(t) &=& \frac{7938}{18608}-\frac{9240}{18608} \cos{\left(\pi\left(1+\frac{t}{\Delta t}\right)\right)} \nonumber\\
&& + \frac{1430}{18608} \cos{\left(2\pi\left(1+\frac{t}{\Delta t}\right)\right)} 
\eeqn
for $|t|<\Delta t$, and $W(t)=0$ otherwise.
The choice of the window size $\Delta t$ is a trade-off between high time resolution (small $\Delta t$) and
high spectral resolution (large $\Delta t$). For Fig.~\ref{fig:spectrogram}, we use $\Delta t=6\,{\rm ms}$, which causes
a noticeable smoothing of the spectrogram in time but is necessary to start resolving the secondary peaks of emission. 

In the spectrograms, the fundamental peak is clearly visible, and strongest for the softest equation of state (SFHo). It is only mildly damped
over the short duration of the simulations, but varies in frequency as the structure of the remnant evolves. This shift
can be as large as $200{\,\rm Hz}$ in the case of the DD2 equation of state, and mostly occurs in the first $\sim 5\,{\rm ms}$ after
merger. Fig.~\ref{fig:spectrogram} also clearly
shows that the gravitational wave emission at the secondary peaks is extremely short-lived, with a decay timescale
of $\sim 1\,{\rm ms}-3\,{\rm ms}$. The emission at the secondary peak is both weaker and shorter-lived for the softest equation
of state (SFHo). For low-mass systems, we thus see that only the fundamental mode remains significantly excited in the
post-merger remnant. Gravitational wave emission at lower frequencies is largely coincident with the merger itself, 
and the secondary peaks are naturally broad in spectral space as the signal decays over only a few oscillation
periods. For the low-mass systems studied here, there is thus a significant difference between the strong, long-lived
peak corresponding to the fundamental mode, and the weak, broad and short-lived peaks at lower frequencies,
which would naturally make the latter difficult to observe or disentangle from detector noise.

\begin{table}
\caption{
  Frequency of the two strongest peaks in the spectrum of the post-merger gravitational wave signal, $f_0$,
  and $f_1$. The prediction from~\cite{Bauswein:2015a} for the dominant peaks, obtained from simulations using an approximate
   treatment of gravity, are listed as $f_{\rm peak}^{\rm fit}$, $f_{\rm spiral}^{\rm fit}$ and $f_{\rm 2-0}^{\rm fit}$. The prediction
   from~\cite{Bernuzzi2015} for $f_0$ is $f_{\rm peak}^{\rm B}$.
}
\label{tab:gw}
\begin{ruledtabular}
\begin{tabular}{|c|c|c|c|c|c|c|}
{\rm EoS} & $f_0$ & $f_1$ & $f_{\rm peak}^{\rm fit}$ &  $f_{\rm spiral}^{\rm fit}$ & $f_{\rm 2-0}^{\rm fit}$ & $f_{\rm peak}^{\rm B}$\\
\hline
SFHo & 2.96kHz & 2.1kHz  & 3.03kHz & 2.1kHz & 1.7kHz & 2.66kHz\\ 
LS220 & 2.56kHz & 1.8kHz  & 2.59kHz & 1.9kHz & 1.4kHz & 2.27kHz\\
DD2 & 2.35kHz & 1.7kHz  & 2.39kHz & 1.8kHz& 1.2kHz & 2.18kHz\\
\end{tabular}
\end{ruledtabular}
\end{table}

The interpretation of the emission of gravitational waves at frequency $f_{\rm spiral}$ as the result of rotating spiral structures within the core of the remnant is partially supported by visualizations of the rest mass density in the equatorial plane 
(see Fig.~\ref{fig:RhoEq}). 
Right after merger, when the emission at $f_{\rm spiral}$ is the strongest, the LS220 and DD2 simulations show
clear spiral arms, including in high-density regions. In the SFHo simulation, the spiral arms are significantly weaker. Later on, 
$3\,{\rm ms}$ after merger, lower density spiral structures remain visible in the LS220 and DD2 simulations, and are stronger
in the latter simulation.
Again, this is in agreement with the observed difference in the damping timescale of the emission at $f_{\rm spiral}$
observed in the spectrograms (Fig.~\ref{fig:spectrogram}). Finally, $10\,{\rm ms}$ after merger, extended spiral structures
remain visible, but they are confined to low-density regions and are unlikely to significantly contribute to gravitational wave
emission. Our results thus appear consistent with the interpretation of $f_{\rm spiral}$ proposed by 
Bauswein \& Stergoulias~\cite{Bauswein:2015a}. 

An alternative universal relation, this time between the strongest peak of the post-merger gravitational wave signal 
and the tidal coupling constant 
$\kappa_2$, has been proposed by Bernuzzi et al.~\cite{Bernuzzi2015}.
Bernuzzi et al.~\cite{Bernuzzi2015} predict
\beq
f_{\rm peak}^{\rm B} = 4.341 \frac{1+0.00167\kappa_2}{1+0.00656\kappa_2}\,{\rm kHz}\,.
\label{eq:fB}
\eeq
We compare $f_{\rm peak}^{\rm B}$ to our results in Table~\ref{tab:gw}. $f_{\rm peak}^{\rm B}$ is systematically $(0.2-0.3)\,{\rm kHz}$
below the peak frequency measured in our simulations. The magnitude of the error is comparable to the scatter observed 
within the simulations presented in Bernuzzi et al.~\cite{Bernuzzi2015}. It is unclear whether the systematic underestimate
of the peak frequency comes from our
use of a hot nuclear theory based equation of state, the application of~(\ref{eq:fB}) to a low mass system, or the intrinsic scatter
around~(\ref{eq:fB}). Bernuzzi et al.~use piecewise polytropic equations of state
with a $\Gamma$-law thermal component for the pressure, and $f_{\rm peak}^{\rm B}$ is only fitted to mergers with
total mass within the interval $[2.45M_\odot,2.9M_\odot]$, so that comparison with our results requires extrapolation of their fitting
formula to lower mass systems.

\begin{figure}
\flushleft
\includegraphics[width=1.03\columnwidth]{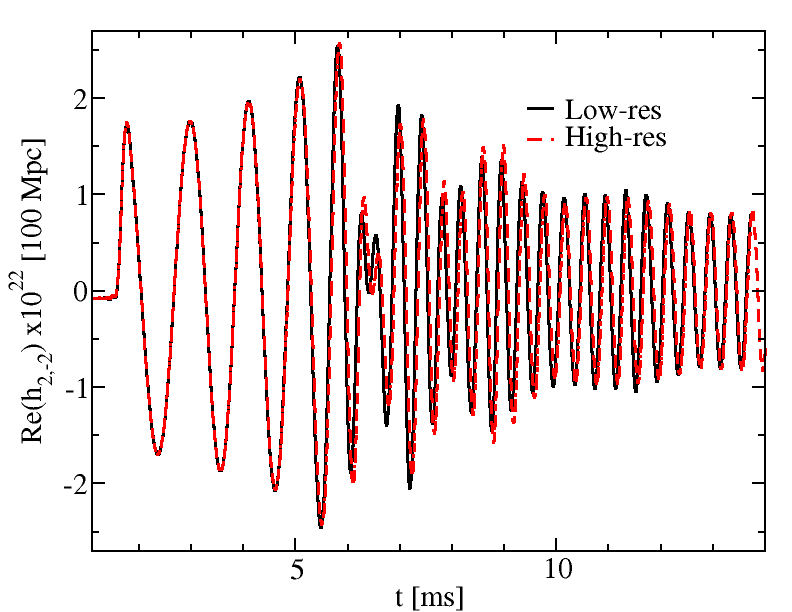}
\caption{Dominant (2,2) mode of the gravitational wave strain for the LS220 waveform at the standard and high resolution,
at $100\,{\rm Mpc}$ from the source.
We find excellent agreement even in the post-merger signal.}
\label{fig:GWconv}
\end{figure}

Comparing the standard and high resolution simulations for the LS220 equation of state does not reveal any significant resolution 
dependence in the spectrum of the gravitational wave signal. Fig.~\ref{fig:spectra} shows mild differences in the shape of the
secondary peaks, which are however also observed when changing the exact interval over which we perform the Fourier transform,
and excellent agreement in the location and amplitude of the primary peak. 
Even in the time domain, the low-resolution and high-resolution
waveforms appear to only differ by a small time shift (see Fig.~\ref{fig:GWconv}). We note that for both 
Fig.~\ref{fig:spectrogram} and Fig.~\ref{fig:GWconv}, we use waveforms extracted at finite radius ($r\sim 130M$) to have
access to a longer post-merger signal. However, Fig.~\ref{fig:spectrogram} shows that only a few milliseconds of post-merger 
signal are actually necessary to study the secondary peaks in the post-merger spectrum, since the gravitational wave emission
at those frequencies decays on a very short timescale.

\section{Merger and Post-Merger Remnant}
\label{sec:remnant}

\begin{figure*}
\includegraphics[width=1.\linewidth]{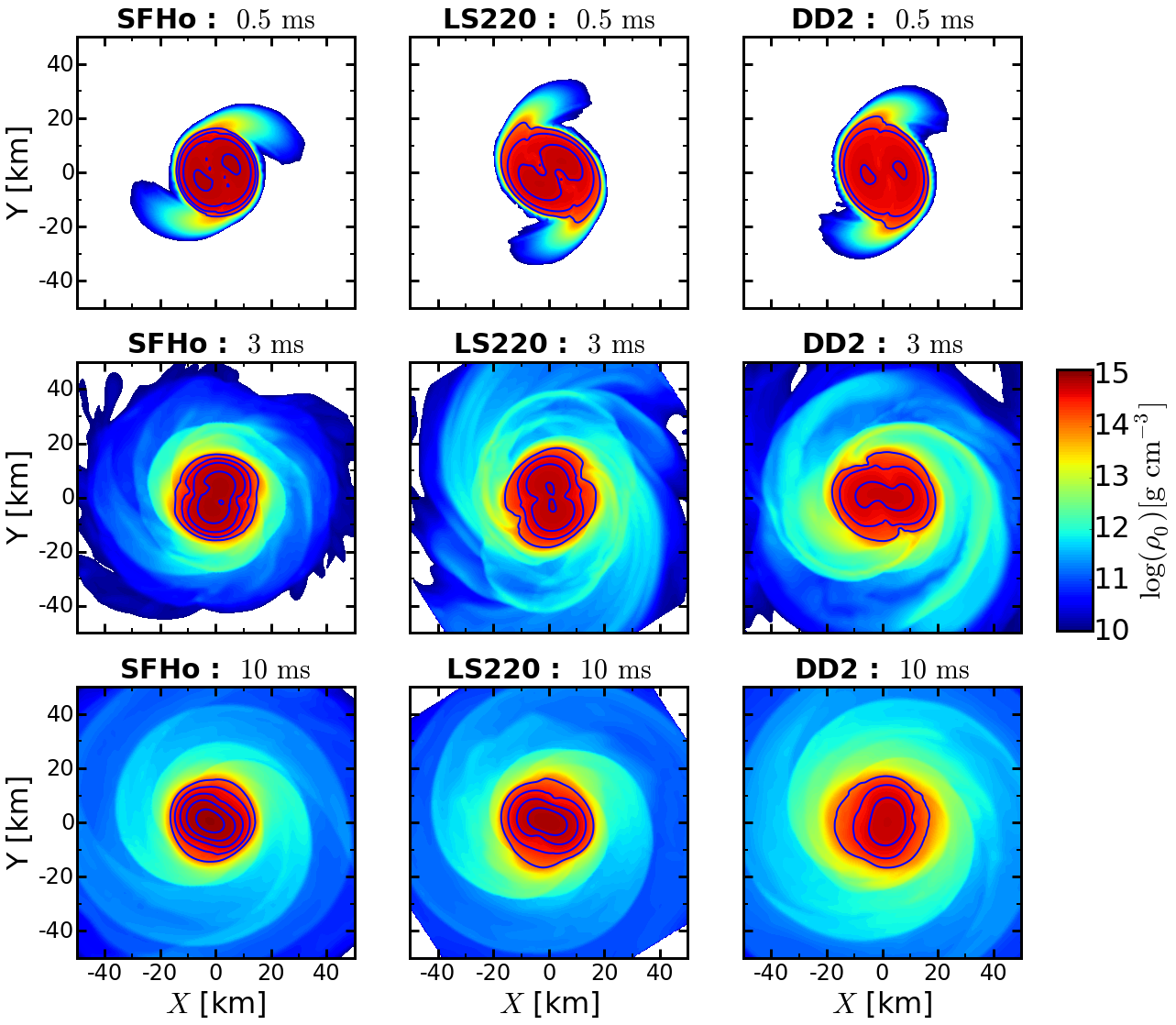}
\caption{Density in the equatorial plane of the post-merger remnant at three characteristic times: $0.5\,{\rm ms}$ after
the peak of the gravitational wave emission (top), $3\,{\rm ms}$ after the peak (middle), and $10\,{\rm ms}$ after the peak
(bottom). We show results for the SFHo (left), LS220 (center) and DD2 (right) equations of state. In each of those plots,
we show density contours at $(1,3,5,7,9)\times 10^{14}\,{\rm g\,cm^{-3}}$ (solid blue lines). Here and in subsequent figures,
the coordinates are those of the simulation, as evolved in the generalized harmonic formulation of Einstein's equations.}
\label{fig:RhoEq}
\end{figure*}

We now consider the qualitative properties of the merger and post-merger remnants. 
Figure~\ref{fig:RhoEq} shows snapshot of the density profile in the equatorial plane of the binary $0.5\,{\rm ms}$, $3\,{\rm ms}$ 
and $10\,{\rm ms}$ after the peak of the gravitational wave amplitude for all 3 equations of state, which we define as the time of 
merger. Initially,
the properties of the system are largely determined by the compactness of the pre-merger neutron stars. With the SFHo
equation of state, i.e. for the most compact neutron stars, a compact core forms rapidly. For the other equations of state (LS220, DD2),
more strongly developed tidal features appear at merger. $0.5\,{\rm ms}$ after merger, we still observe two well-defined cores whose
size and tidal distortion directly correlates to the pre-merger size of the individual neutron stars. 

\begin{figure}
\includegraphics[width=1.03\linewidth]{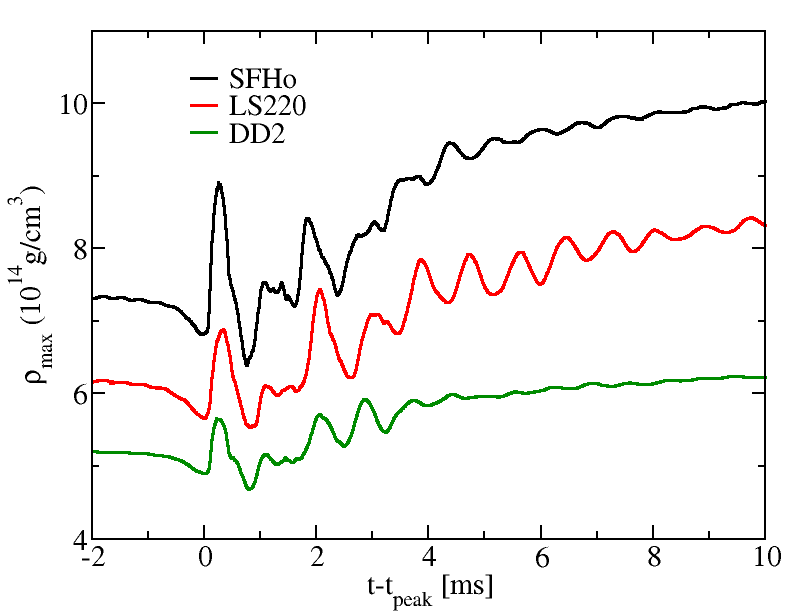}
\caption{Maximum value of the baryon density on the grid as a function of time for all three equations of state. $t_{\rm peak}$
is the time at which the amplitude of the gravitational wave signal is maximal}
\label{fig:MaxRho}
\end{figure}

\begin{figure}
\includegraphics[width=1.\linewidth]{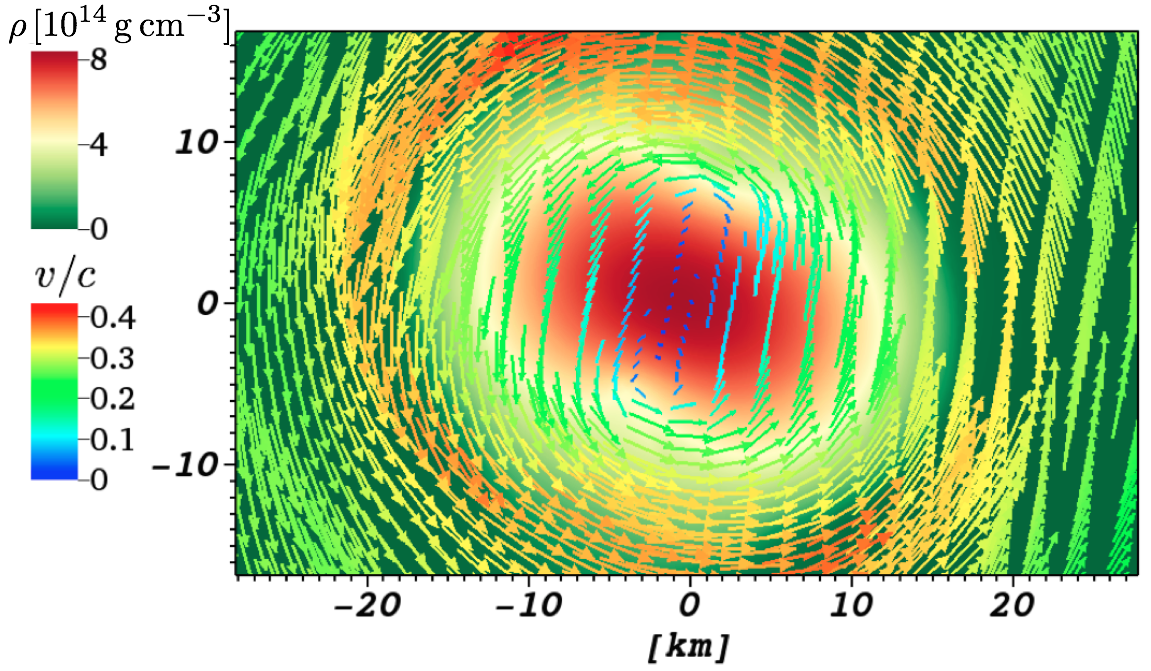}
\caption{Density and velocity in the equatorial plane for the simulation with the LS220 equation of state, $10\,{\rm ms}$
after merger. The excited quadrupole mode is still clearly visible at that time.}
\label{fig:FinalRhoAndV}
\end{figure}

After $3\,{\rm ms}$, the cores
start to merge. The less compact neutron stars (LS220, DD2) produce clearly defined, high density tidal tails. Finally, $10\,{\rm ms}$
after merger, the properties of the remnant are dominated by the high-density behavior of the equation of state. For high
neutron star masses, the LS220 equation of state is nearly as compact as the SFHo equation of state - and neither of them is
able to support a stable, non-rotating, cold $2.4M_\odot$ neutron star. The LS220 and SFHo equations of state now have
similarly compact cores, with central density rising slowly as the neutron star evolves. Since both post-merger remnants have baryon
masses well below the maximum mass of a uniformly rotating cold neutron star (at the mass-shedding limit), 
we expect the post-merger neutron star
to survive for much longer than the duration of the simulation. Collapse to a black hole should occur on the timescale
necessary for the post-merger neutron star to loose angular momentum through magnetically-driven spin-down. 
On the other hand, the DD2 equation of state forms a massive neutron star which will remain stable even in the non-rotating
limit. We see that the post-merger
neutron star evolves only slowly
at late times (see Fig.~\ref{fig:MaxRho}). All three configurations remain far from axisymmetry. Strong spiral waves are
driven in the forming accretion disk, and a rapidly rotating bar mode remains present in the post-merger neutron star 
(see Fig.~\ref{fig:FinalRhoAndV}). 
Figure~\ref{fig:MaxRho} also shows that for all three configurations, the peak of the gravitational wave emission is coincident with the 
core bounce of the merging neutron stars, i.e. the time at which the two cores touch. The core bounce causes a sharp increase 
in the density and pressure of the cores,
as well as strong shocks, shears, and heating at the interface between the merging neutron stars.

\begin{figure*}
\includegraphics[width=1.\linewidth]{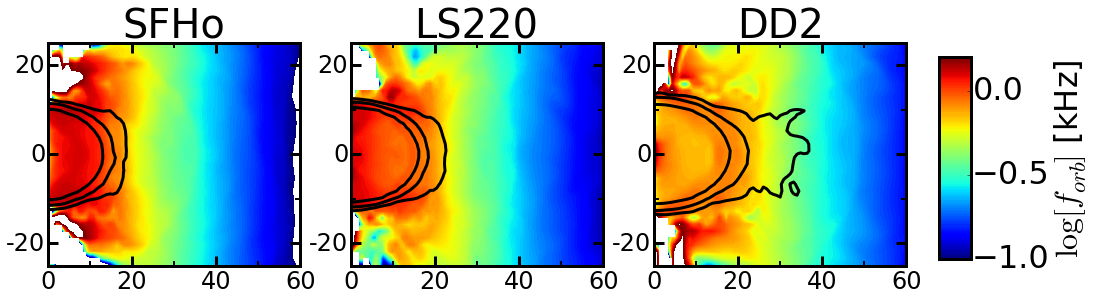}
\caption{Instantaneous orbital frequency of the fluid, in the coordinates of the simulation and $10\,{\rm ms}$ after merrger, 
in a slice orthogonal
to the orbital plane and along the major axis of the bar in the post-merger neutron star. The black lines are density contours
at $\rho_0=(10^{12},10^{13},10^{14})\,{\rm g/cm^3}$.}
\label{fig:OmegaBar}
\end{figure*}

The excitation of the post-merger remnant, shown in Fig.~\ref{fig:FinalRhoAndV}, is consistent with the fundamental
quadrupole mode, both in terms of the density and velocity patterns, and in terms of the measured gravitational
wave frequency. Because of this excitation, the spatial variation of the instantaneous orbital angular velocity, defined
in the coordinates of the simulation and with respect to the center of mass of the original binary, is fairly complex. We
show the instantaneous orbital frequency of the fluid in Fig.~\ref{fig:OmegaBar}. Although on average the rotation frequency
is higher at smaller radii, we observe regions in which $d\Omega/dr>0$.
There are thus regions of the post-merger neutron star which may not be subject to the axisymmetric
magnetorotational instability (MRI)~\cite{BalbusHawley1991}, at least shortly after merger. 
However, all regions with densities $\rho_0\lesssim 10^{13}\,{\rm g\,cm^{-3}}$ follow a 
typical disk profile for the orbital frequency, and should see rapid growth of the magnetic field. The typical timescale for
the growth of the axisymmetric MRI is $\tau_{\rm MRI} \sim 4/(3\Omega) \sim 1\,{\rm ms}$~\cite{BalbusHawley1991}. 
Magnetic field should thus
grow significantly in the disk over even the short timescale of our simulation. So far, simulations have
only been able to resolve the growth of the MRI in lower-density regions 
($\lesssim 10^{12}\,{\rm g/cm^3}$)~\cite{Kiuchi2014}, and indeed find rapid growth of the magnetic field at early times
due to a combination of magnetic instabilities and winding of the magnetic field.
We should also note that the angular velocity in the remnant varies significantly when comparing a vertical slice along the
direction of the bar mode in the remnant (as in Fig.~\ref{fig:OmegaBar}), with a slice orthogonal to that direction (see, e.g.,
the azimuthal dependence of the velocity on Fig.~\ref{fig:FinalRhoAndV}). Figure~\ref{fig:OmegaBar} may give the impression
that the bar mode is rotating at frequencies much lower than the expected $f_{\rm peak}/2$. In fact, we can check
that the pattern speed of the bar mode is consistent with $f_{\rm peak}/2$. In the LS220 simulation, we measure a rotation of the 
bar of $\sim 45^\circ$ over the last $0.1\,{\rm ms}$ of evolution, or a frequency $f_{\rm bar}\sim 1.25\,{\rm kHz}\sim f_{\rm peak}/2$.
We also find that the orbital frequency in the core of the post-merger neutron star is higher than the mass-shedding limit for a uniformly rotating neutron
star of the same baryon mass. Considering that the total baryon mass of the system is below the maximum mass of a cold, uniformly
rotating neutron star (at the mass-shedding limit), this suggests that rotational support is initially sufficient to prevent
the collapse of the remnant to a black hole. This should be true even if the differential rotation is rapidly erased by angular momentum
transport within the remnant.

\begin{figure*}
\includegraphics[width=1.\linewidth]{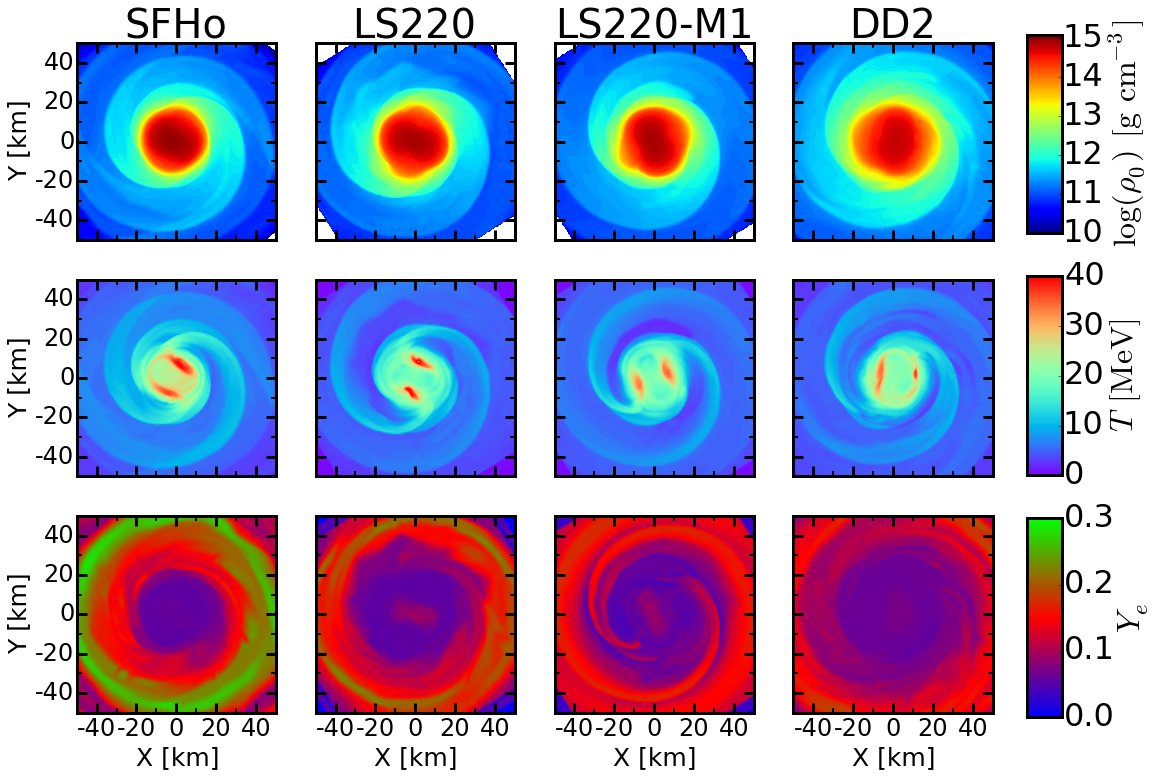}
\caption{Hydrodynamic variables (in the equatorial plane) $10\,{\rm ms}$ after the peak of the gravitational wave signal.
From top to bottom, we show the density, temperature, and electron fraction for the four simulations 
considered in this paper -- from left to right, SFHo, LS220 with neutrino leakage, LS220 with neutrino transport, and DD2.}
\label{fig:EqSlice}
\end{figure*}

\begin{figure*}
\includegraphics[width=1.\linewidth]{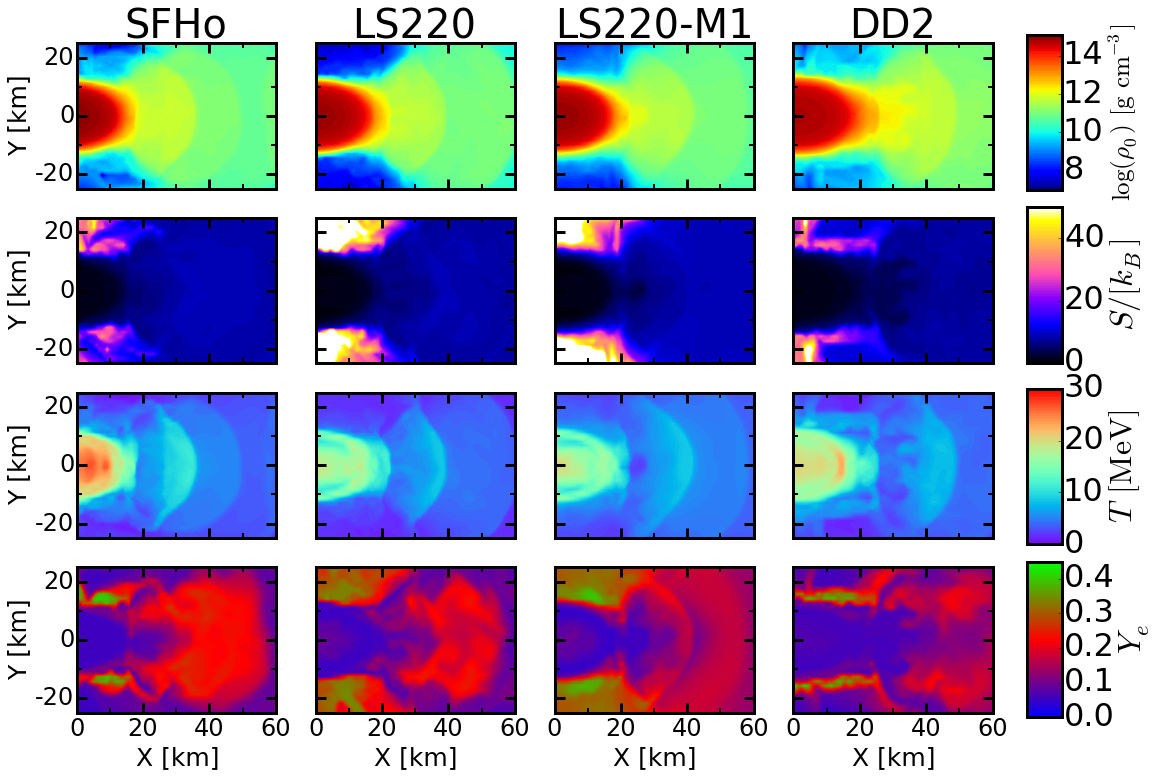}
\caption{Hydrodynamic variables $10\,{\rm ms}$ after the peak of the gravitational wave signal in a slice along the major 
axis of the remnant and orthogonal to the equatorial plane.
From top to bottom, we show the density, entropy per baryon, temperature, and electron fraction for the four simulations 
considered in this paper -- from left to right, SFHo, LS220 with neutrino leakage, LS220 with neutrino transport, and DD2.}
\label{fig:BarSlice}
\end{figure*}

The properties of the post-merger remnant are strongly affected by the presence
of a non-linear $l=2$ perturbation. Visible shock fronts remain in the disk at the end of the simulation,
while a clear bar mode is observed in the post-merger neutron star.
In Fig.~\ref{fig:EqSlice} and Fig.~\ref{fig:BarSlice}, we plot the density, temperature, composition and entropy of the fluid 
in the equatorial plane and in a meridional slice along the major axis of the bar observed in the post-merger neutron star. For all
three simulations, the neutron star remains neutron rich and, although hotter than the disk in terms of absolute temperature, 
with $T\sim(15-25)\,{\rm MeV}$, remains entirely supported by nuclear forces and rotation~\cite{Kaplan2013}. The entropy
per baryon in the neutron star is only $S\sim 1k_B$. The neutron star is thus colder than the $30-50\,{\rm MeV}$
temperatures observed in higher mass systems~\cite{Hotokezaka2013}, presumably due to the fact that lower
mass stars are less compact, and thus that shock heating of the neutron star during the merger is not as important of an effect
in the low-mass systems considered here.
This interpretation is partially confirmed by the fact that the softer equation of state (SFHo) has a core temperature
significantly above the stiffer equations of state (DD2, LS220). All three simulations however show significant heating
at the interface between the bar and the disk, with $T\sim(30-40)\,{\rm MeV}$. At the lower densities observed in that
region, the thermal pressure now becomes important.

The maximum density in the accretion disk remains high, $\rho_{\rm disk}\gtrsim 10^{12}\,{\rm g\,cm^{-3}}$. 
The shock front associated with the $l=2$ perturbation of the disk heats the disk, causing temperatures of $(8-10)\,{\rm MeV}$ right at the shock.
Accordingly, the disk is thick, with a scale height $H\sim R$, and thermally supported, with $P(\rho_0,T,Y_e)\gg P(\rho_0,0,Y_e)$:
for $\rho=10^{12}\,{\rm g\,cm^{-3}}$, $P(\rho_0,T,Y_e)\approx 2 P(\rho_0,0,Y_e)$ for $T=2\,{\rm MeV}$, and most of the disk is
much hotter than that. 

We test the post-merger stars and disks for axisymmetric convective
instability using the relativistic Solberg-Hoiland-Ledoux
condition~(\cite{1975ApJ...197..745S} generalized
to include $Y_e$ gradient terms).  This condition assumes an
axisymmetric, stationary
background, which is only approximately present, so we use density-weighted
azimuthal average profiles.  Consider the LS220 case, which is evolved both
in leakage and M1.  Concentrating on the region near the equator, where
the radial condition is most important, we find that the star and disk
are stable everywhere except for a small region around cylindrical radius
of 55\,km, where the angular momentum gradient is Rayleigh unstable, an
indication that matter at this distance has not yet acheived equilibrium. 
The star and bulk of the disk are convectively stable even though the
$Y_e$ gradient is unstable through much of the star.  This effect is
counteracted by the much larger stabilizing shear and entropy gradient terms. 

We can compare our results with recent simulations using nuclear-theory based equations of state at similar compactness and
higher mass~\cite{Neilsen:2014hha}, similar radii and higher mass~\cite{Palenzuela2015,Sekiguchi:2015}, and 
simulations approximating the thermal  dependence of the equation of state by a $\Gamma$-law for a wide range of systems 
of higher mass~\cite{Hotokezaka2013}, and a single system with a large neutron star of similar 
mass~\cite{Giacomazzo:2013uua}. 
The temperature of the remnant and tidal features are closest to those found
by Neilsen et al.~\cite{Neilsen:2014hha}, confirming the correlation of those features with the compactness of the neutron star. 
More compact neutron stars generally show stronger shock heating in the post-merger neutron star, stronger excitation
of the bar mode in the core, and weaker $l=2$ features in the disk. The first two are
due to smaller neutron stars merging at closer separation and thus higher velocities, causing a more violent merger event.
The third is presumably due to stronger tidal effects in less compact stars. The main features of the merger in our simulations are in fact
remarkably similar to those observed in~\cite{Neilsen:2014hha,Palenzuela2015}, especially when one takes into
account the expected dependence of the results on the compactness of the star. The agreement between our results 
and~\cite{Neilsen:2014hha,Palenzuela2015} is reassuring considering the use of similar equations
of state and a mostly identical neutrino leakage scheme. 

\begin{figure}
\includegraphics[width=1.\linewidth]{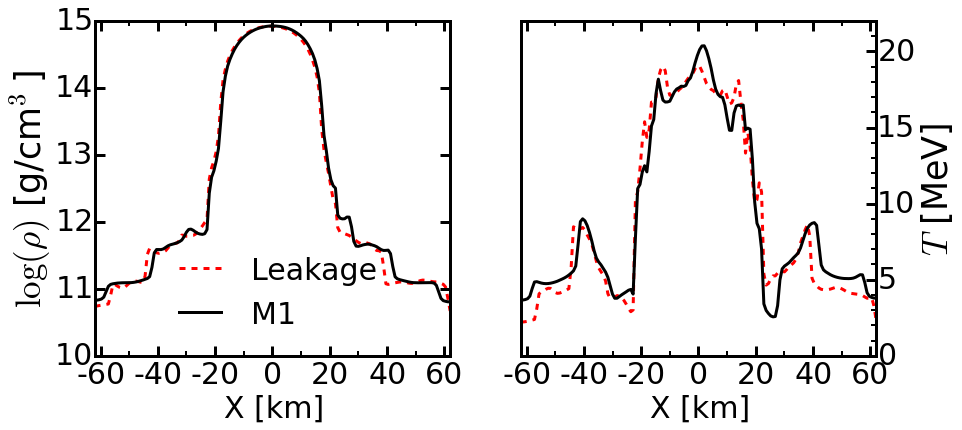}
\caption{Density ({\it left}) and temperature ({\it right}) profiles in the equatorial plane along the major axis of the merger remnant.
We show results for the merger using the LS220 equation of state and the neutrino transport (solid lines) or neutrino leakage (dashed
line) scheme. The two simulations are very similar, with a slightly smoother temperature profile and hotter outer disk when using
neutrino transport. The hot shocked tidal arms are clearly visible around $40\,{\rm km}$ from the center in both plots.}
\label{fig:1Dprof}
\end{figure}

This agreement does not, however, preclude systematic errors due to the limited physics included in the simulations. One
such limitation is the use of a leakage scheme to treat neutrino cooling, and the absence of any treatment of neutrino
absorption. We can test the effect of these assumptions by using a neutrino transport scheme. To do
this, we evolved the merger with the LS220 equation of state using our M1 transport scheme, which evolves the energy
density and momentum density of the neutrinos~\cite{Foucart:2015a}. Figures~\ref{fig:EqSlice}-\ref{fig:BarSlice} show that the 
qualitative evolution of the neutron star remnant is largely unaffected by the treatment of the neutrinos. This is not at all surprising,
given that at the large densities existing within the neutron star, the neutrinos are trapped and in equilibrium with the fluid (see also 
Sec.~\ref{sec:neutrinos}). As already observed in the disks resulting from NSBH mergers~\cite{Foucart:2015a}, the inclusion of neutrino 
transport causes a smoothing of the temperature profile. This is particularly visible in the equatorial plane at the bar-disk
interface. Neutrino absorption also naturally causes cold 
low-density regions in the disk to be heated: most of the disk is $\sim 1\,{\rm MeV}$ hotter when using 
neutrino transport. The mild smoothing of the temperature profile, heating of the outer disk, and shocked tidal arms
can also be observed in the one-dimensional density and temperature profiles presented on Fig.~\ref{fig:1Dprof}.

These are fairly minor effects, with some impact on the neutrino luminosity and neutrino-matter interactions (see Sec.~\ref{sec:neutrinos}),
but not on the hydrodynamic properties of the post-merger remnant. The treatment of the neutrinos really begins to matter when one considers the
composition of the disk, as measured by its electron fraction $Y_e = n_p/(n_p+n_n)$, with $n_p$ and $n_n$ the
number density of protons and neutrons respectively. In leakage simulations the neutrino emission, and thus
the composition of the disk, is set only by the local properties of the fluid and an estimated neutrino optical depth. 
The transport scheme, on the other hand, takes into account the irradiation of cold regions of the disk by the hot neutrino-emitting regions. 
Most of the neutrino emission occurs either at the core-disk interface or at the tidal
shocks in the disk, and the composition of the disk is largely set by the relative position of a fluid element with respect to these
two defining features. We will see in Sec.~\ref{sec:neutrinos} that the disk is mostly irradiated by electron antineutrinos.
Accordingly, the regions immediately adjacent to the neutrino-emitting regions, i.e. just next to the hot shock front in the disk
or the surface of the post-merger neutron star, are forced towards a very low
electron fraction $Y_e\sim 0.05-0.1$. Farther away from those emitting regions, we get $Y_e\sim 0.15-0.2$.
The leakage scheme, which does not take into account the absorption of electron antineutrinos in the disk, predicts
an electron fraction $Y_e\sim 0.2-0.3$ in most of the high-density regions of the disk, with smaller $Y_e$ in the corona
and at low radii. There is thus a very significant difference in the composition of the disk between leakage and transport
simulations, due to the fact that the composition of most of the disk is strongly affected by neutrino absorption. This can
be contrasted with the remnant of NSBH mergers, where only the low-density corona
is significantly affected by neutrino absorption~\cite{Foucart:2015a}

We also note that neutrino absorption in the transport simulation causes the shocked regions in the tidal arms of the disk 
to be about $0.5\,{\rm MeV}$ hotter than predicted by the leakage scheme (see Fig.~\ref{fig:1Dprof}). The shock front
is energized by neutrinos emitted in the inner disk or in the core. 

Despite these effects,
our simulations indicate that neutrino leakage is a reasonable approximation as long as the exact composition of the fluid 
is not required, i.e. except when attempting to make predictions for r-process nucleosynthesis in the dynamical ejecta and
disk winds, and for the associated radioactively powered electromagnetic signal.

Finally, we note that for all the variables discussed in this section, the low and high resolution simulations with the LS220
equation of state give results well below the differences between simulations using different equations of state. 
We thus expect the results to be only
mildly affected by the numerical error in the simulations.

\section{Neutrino Emission}
\label{sec:neutrinos}

\begin{figure}
\flushleft
\includegraphics[width=1.03\columnwidth]{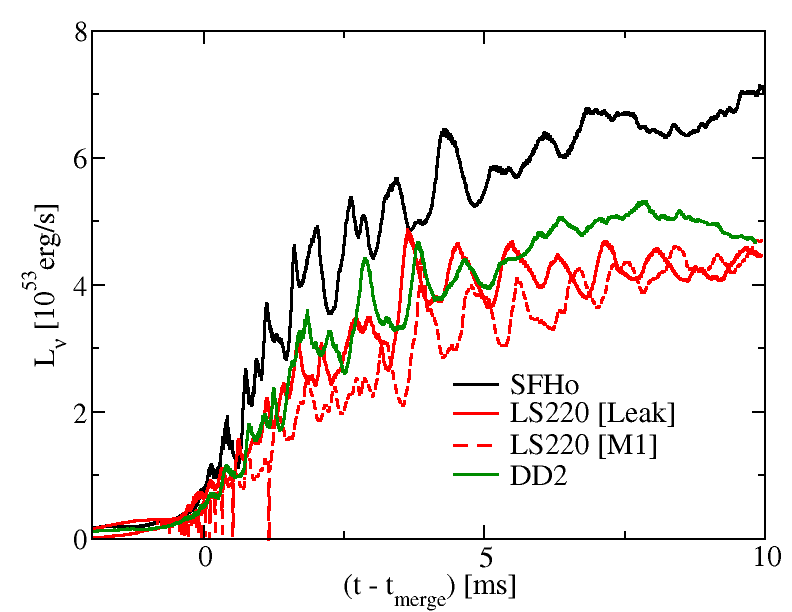}
\caption{Total neutrino luminosity for the three simulations using the leakage scheme and the LS220, SFHo, and DD2 
equations of state (solid curves)
as well as the simulation using the LS220 equation of state and neutrino transport (dashed red curve).}
\label{fig:totallum}
\end{figure}

\begin{figure}
\flushleft
\includegraphics[width=1.03\columnwidth]{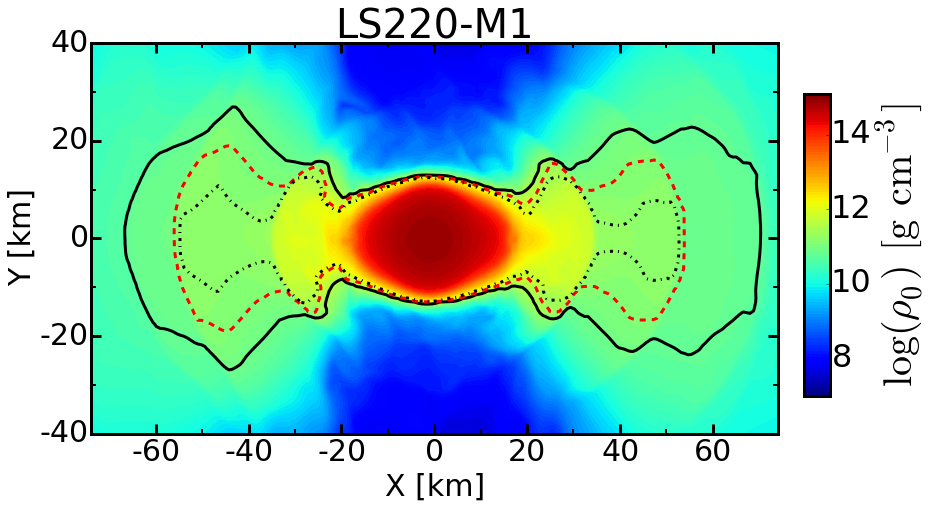}
\caption{Density (colour scale) and location of the surfaces of optical depth $\tau\sim 1$ for $\nu_e$ (solid black line), $\bar\nu_e$
(dashed red line), and $\nu_x$ (dot-dashed black line) at the end of the simulation using the LS220 equation of state and 
a neutrino transport scheme.}
\label{fig:neutrinospheres}
\end{figure}

When studying the neutrino emission in simulations using the leakage scheme, we can consider only the total
energy and number of emitted neutrinos, as well as the predicted location of emission. We list the luminosity and average energy
of neutrinos in all three simulations in Table~\ref{tab:nu}, and provide the time dependence of the total luminosity 
in Fig.~\ref{fig:totallum}. For the average energy, we compute the energy-weighted root-mean-square energy, which is the 
important quantity when estimating the energy deposited in optically thin regions through neutrino absorption. 
For a blackbody spectrum, as assumed here, it is a factor of $\sim 1.5$ larger than the number-weighted average energy of the neutrinos. We find results similar to simulations of higher mass neutron star mergers using
leakage schemes~\cite{Neilsen:2014hha,Palenzuela2015} 
and leakage-transport hybrid~\cite{Sekiguchi:2015} schemes. The electron antineutrinos have the highest luminosity,
with $L_{\bar\nu_e}\sim (2-3) \times 10^{53}\,{\rm erg\,s^{-1}}$ about $10\,{\rm ms}$ after merger. They dominate the
electron neutrinos by a factor of $1.4-2$. 

To understand the geometry of the neutrino radiation, it is useful to determine which regions of the remnant are optically thick
to neutrinos, and which are not. To do so, we show on Fig.~\ref{fig:neutrinospheres} the location of the surface of optical depth
$\tau\sim1$ for all species of neutrinos, as computed by the leakage scheme. 
The remnant neutron star is naturally optically thick to all neutrinos. For the electron neutrinos
and antineutrinos, this is also the case for most of the inner disk, up to or even further than the location of the shocked tidal arms.

The electron neutrinos and antineutrinos are emitted from different regions of
the remnant (see also~\cite{Neilsen:2014hha}): as will be made clearer when considering the results of the simulation
using neutrino transport, most electron neutrinos come from the polar regions of the core of the remnant, 
but most electron antineutrinos
are emitted in the hot shocked regions of the disk. This is largely due to the fact that these shocked regions are surrounded
by material which has low optical depth to $\bar\nu_e$, but is optically thick to $\nu_e$ (see Fig.~\ref{fig:neutrinospheres}). 
The heavy
lepton neutrinos, whose emissivity is more sensitive to the temperature of the fluid, are mostly emitted in the hottest regions
of the core. Accordingly, the luminosity in $\bar\nu_e$ scales with the temperature of the shocked regions of the disk while the
luminosity in $\nu_e$ and heavy-lepton neutrinos scales with the temperature of their respective neutrinosphere in the core
(i.e.~the region of optical depth of order unity, which is closer to the surface for $\nu_e$ than for the heavy-lepton neutrinos). 
For all species, the highest luminosity is naturally reached in the case of the most compact neutron stars (SFHo equation of state), 
given the stronger heating of the remnant occurring
in that case. The LS220 equation of state, which forms a relatively cold core (see e.g.~Fig.~\ref{fig:BarSlice}), 
shows a particularly large difference between
the emission of electron antineutrinos and electron neutrinos.

The average energy of the electron antineutrinos is set by the equilibrium spectrum in the
shocked regions of the disk, since they are emitted there and then largely free-stream away from the disk. 
The average energy of the other species is set by the temperature of the core at the point
at which the neutrinos decouple from the fluid, which occurs at higher density (and temperature) 
for the heavy-lepton neutrinos than for the electron neutrinos. 

\begin{table}
\caption{
  Neutrino luminosities and rms energy of the neutrinos according to the leakage scheme, $10{\rm ms}$
  after merger. The $\nu_x$ values are for all heavy-lepton neutrinos and antineutrinos combined.
}
\label{tab:nu}
\begin{ruledtabular}
\begin{tabular}{|c|ccc|ccc|}
{\rm EoS} & $L_{\nu_e}$ & $L_{\bar\nu_e}$ & $L_{\nu_x}$ &  $\sqrt{\langle \epsilon_{\nu_e}^2 \rangle}$ & $\sqrt{\langle \epsilon_{\bar\nu_e}^2 \rangle}$ & $\sqrt{\langle \epsilon_{\nu_x}^2 \rangle}$\\
Units &  & $10^{53}{\rm erg\,s^{-1}}$ &  & & MeV  &  \\
\hline
SFHo & 1.9 & 3.0 & 2.2 & 14 & 21 & 29\\ 
LS220 & 1.2 & 2.1 & 1.2 & 13 & 20 & 26 \\
DD2 & 1.6 & 2.2 & 0.9 & 13 & 20 & 24\\
\end{tabular}
\end{ruledtabular}
\end{table}

At the high luminosities observed here, we expect neutrino absorption in low-density regions to be important to the 
evolution of the composition of the disk and of the outflows. The geometrical properties of the neutrino distribution and the
impact of the neutrinos on the evolution of the system may not be suitably captured by the leakage scheme.
We estimate the impact of the use of the approximate leakage scheme by considering the results of the simulation using the
neutrino transport scheme. In that simulation, we evolve the neutrino energy and momentum density on our numerical grid,
and take into account neutrino absorption and scattering in low-density regions. 

\begin{figure}
\flushleft
\includegraphics[width=1.\columnwidth]{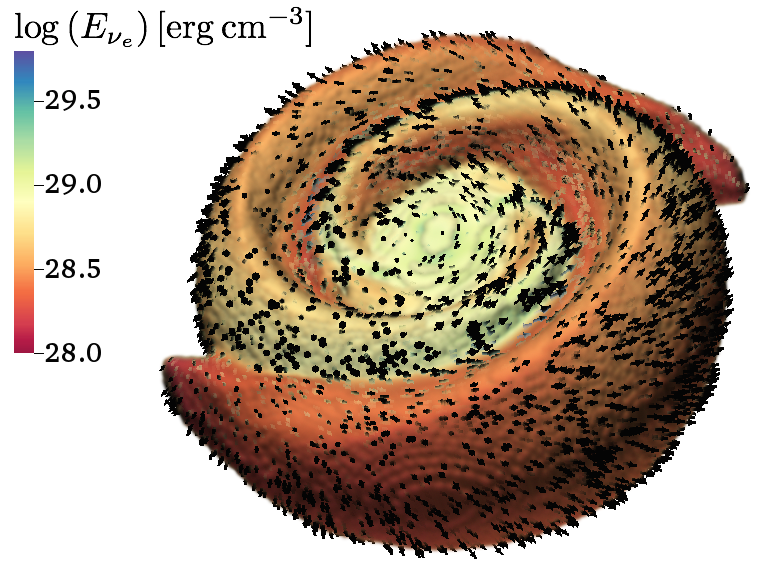}
\caption{Energy density (color scale) and normalized flux $\alpha F^i_\nu/E_\nu-\beta^i$ (black arrows) of the electron neutrinos on 
the density isosurface 
with $\rho_0=10^{11}\,{\rm g/cm^3}$. The region of brightest $\nu_e$ emission, at the center of the plot, coincides with the polar
region of the neutron star remnant.}
\label{fig:enue}
\end{figure}

\begin{figure}
\flushleft
\includegraphics[width=1.\columnwidth]{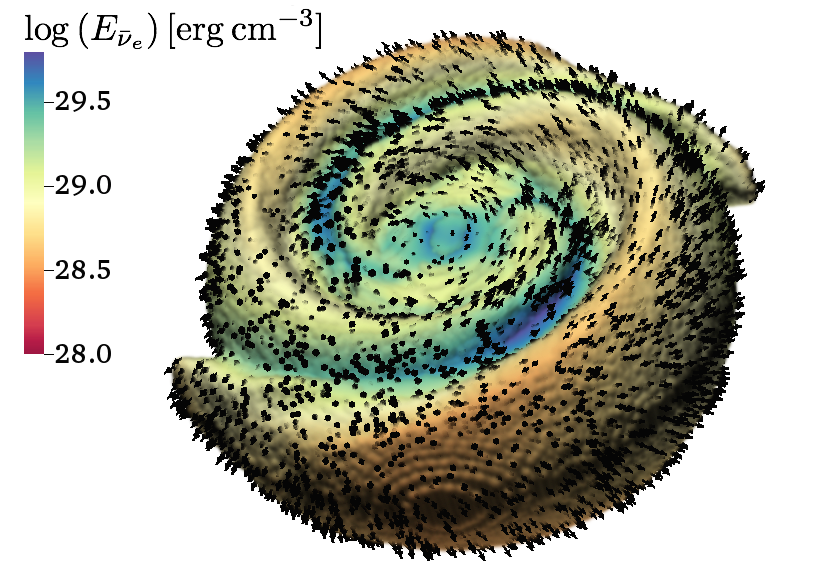}
\caption{Energy density (color scale) and normalized flux $\alpha F_\nu^i/E_\nu-\beta^i$ (black arrows) of the electron 
antineutrinos on the density 
isosurface with $\rho_0=10^{11}\,{\rm g/cm^3}$. The regions of brightest $\bar\nu_e$ emission (dark blue) are in the hot, shocked
tidal arms.}
\label{fig:enua}
\end{figure}

Figures~\ref{fig:enue} and~\ref{fig:enua} show the energy density $E_\nu$ of electron neutrinos and antineutrinos on the surface of matter
density $\rho_0 = 10^{11}\,{\rm g/cm^3}$, together with the neutrino "transport" velocity 
$v_{T,\nu}^i = \alpha F^i_\nu/E_\nu -\beta^i$, i.e.~the velocity such that in the absence of source terms 
$\partial_t (\sqrt{g}E_\nu) + \partial_i(\sqrt{g}E_\nu v_{T,\nu}^i)=0$, and the energy density $E_\nu$ is simply transported
through the grid at velocity $v_{T,\nu}^i$. Here, $F_\nu$ is the neutrino momentum density. We confirm that most of the emission of 
$\bar \nu_e$ comes from the
tidal arms in the disk (dark blue regions in Fig.~\ref{fig:enua}), while most $\nu_e$ escaping the system are emitted in the polar regions of the massive neutron star remnant (center of Fig.~\ref{fig:enue}). We note, however, that even in the polar regions
the electron neutrinos are outnumbered by the electron antineutrinos. Emission from the tidal arms is significantly beamed, due to fluid velocities $v\sim 0.3c$. The emission coming from the polar regions has a wider opening angle.

\begin{figure}
\flushleft
\includegraphics[width=1.\columnwidth]{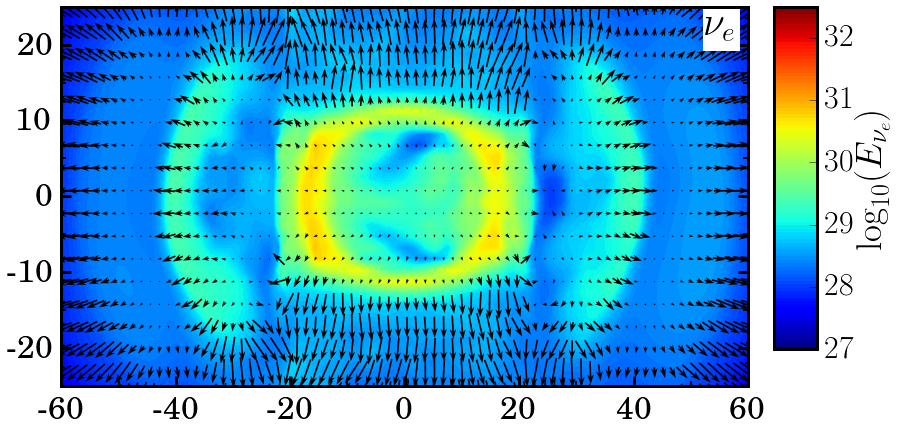}
\caption{Energy density and normalized flux $\alpha F_\nu^i/E_\nu-\beta^i$ of the electron neutrinos in the same vertical
slice as in Fig.~\ref{fig:BarSlice}.
The vertical stripes with higher neutrino energy in the polar regions are an artifact of the M1 closure.}
\label{fig:slicenue}
\end{figure}

\begin{figure}
\flushleft
\includegraphics[width=1.\columnwidth]{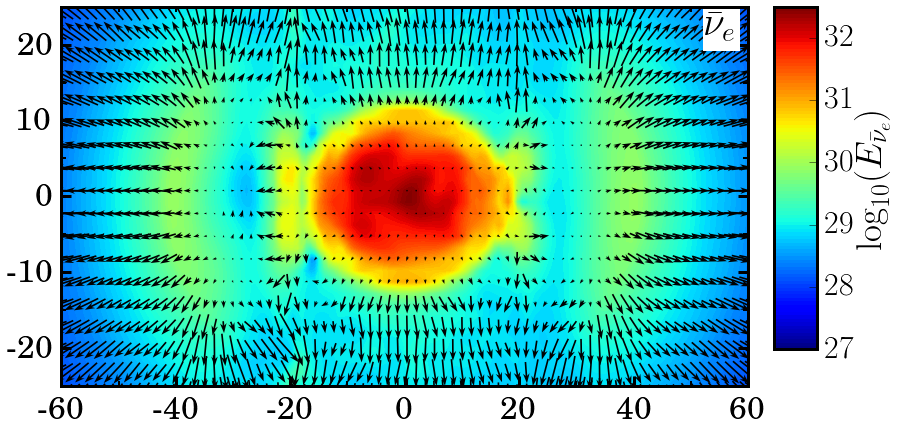}
\caption{Energy density and normalized flux $\alpha F_\nu^i/E_\nu-\beta^i$ of the electron antineutrinos in the same vertical
slice as in Fig.~\ref{fig:BarSlice}.
The vertical stripes with higher neutrino energy in the polar regions are an artifact of the M1 closure.}
\label{fig:slicenua}
\end{figure}

\begin{figure}
\flushleft
\includegraphics[width=1.\columnwidth]{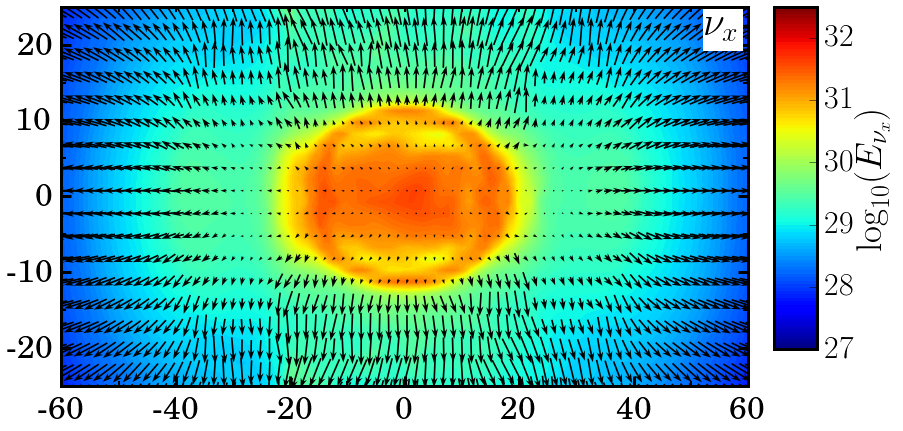}
\caption{Energy density and normalized flux $\alpha F_\nu^i/E_\nu-\beta^i$ of the heavy-lepton neutrinos in the same vertical
slice as in Fig.~\ref{fig:BarSlice}.}
\label{fig:slicenux}
\end{figure}

Figures~\ref{fig:slicenue} to~\ref{fig:slicenux} offer a different view of the same variables, through slices orthogonal to the
equatorial plane and parallel to the major axis of the bar in the remnant. The differences between the electron neutrinos,
the electron antineutrinos, and the heavy-lepton neutrinos are clearly visible.
From the small fluxes seen in Fig.~\ref{fig:slicenue}, we can see that the electron neutrinos are trapped not only in the 
post-merger neutron star, but also in most of the disk -- up to the second, weaker tidal shock at $r\sim 55\,{\rm km}$ (also
visible on Fig.~\ref{fig:BarSlice}). 
Electron neutrinos mostly escape through the polar regions. Most of the electron neutrinos emitted by the hot central 
neutron star toward the disk will be absorbed in the disk, due to disk optical depth $\tau \gg 1$ (see Fig.~\ref{fig:neutrinospheres}). 
The optical depth of the fluid to electron antineutrinos is smaller. Only the regions inside of the first tidal shock
in the disk are optically thick, as visible from the large radial fluxes observed at radii $r\gtrsim40\,{\rm km}$ in 
Fig.~\ref{fig:slicenua} (but not in Fig.~\ref{fig:slicenue}).
 Electron antineutrinos emitted from the shocked regions at $r\sim40\,{\rm km}$ are nearly free-streaming through the outer disk. 
Nearly all heavy-lepton neutrinos are emitted in the hot regions of the core, and are free-streaming
in most of the disk. Finally, we note that the core of the remnant is more neutron rich than the equilibrium value at the temperature of the core ($Y_e^{\rm eq}\sim 0.11-0.12$ at the center of the remnant, where we observe $Y_e\sim 0.05-0.1$). 
This results on a negative equilibrium electron neutrino chemical potential, and leads to a suppression of the electron neutrino density compared to the electron antineutrino density in the core. 
This is opposite to the situation found in the cores of protoneutron stars in a core-collapse supernova (see e.g.~\cite{Roberts2012}).

Figures~\ref{fig:slicenue} and~\ref{fig:slicenua} also show a known limitation of the M1 transport
scheme: its inability to handle crossing beams (see e.g.~the code tests presented in~\cite{Foucart:2015a}). Artificial shock fronts develop in the neutrino evolution in the regions
in which neutrinos from the disk and from the core converge. There, neutrinos coming from different directions start to 
propagate in their average direction of motion instead of crossing each others. This
reduces the opening angle of electron neutrinos and antineutrinos leaving through the polar regions, and could plausibly
affect the composition of matter outflows in those regions.

\begin{figure}
\flushleft
\includegraphics[width=1.\columnwidth]{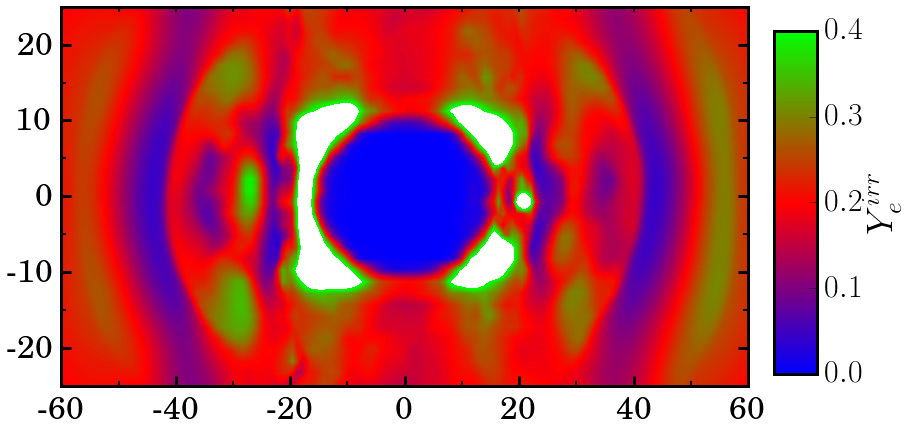}
\caption{Expected equilibrium electron fraction of the fluid $Y_e^{\rm irr}$ if the composition was set solely by 
neutrino absorption. The white region has $Y_e^{\rm irr}>0.4$.}
\label{fig:YeIrr}
\end{figure}

From these observations, we can understand better the differences between the leakage and transport results discussed
in Sec.~\ref{sec:remnant}. In
the post-merger neutron star and in the inner parts of the disk, neutrinos are either in equilibrium with the dense matter or
slowly diffusing through it without much of an effect on its composition over the short timescales
considered here. Accordingly, transport and leakage results are very similar. In the outer disk 
and in the polar regions, low-density material is strongly irradiated by neutrinos. 
If neutrino irradiation was the only driver of composition evolution (i.e.~excluding neutrino emissions and the advection of trapped neutrinos), 
we would expect the fluid to have composition (assuming absorption cross-sections proportional to $\epsilon_\nu^2$)
\beq
Y_e^{\rm irr} \sim \frac{\langle \epsilon_{\nu_e}\rangle E_{\nu_e}}{\langle \epsilon_{\nu_e}\rangle E_{\nu_e}+\langle \epsilon_{\bar\nu_e}\rangle E_{\bar\nu_e}},
\eeq
a quantity plotted in Fig.~\ref{fig:YeIrr} for our simulation. Otherwise, the effect of absorption is at least to
push the composition closer to $Y_e^{\rm irr}$ than what one would expect when neglecting absorption.

The impact of neutrino absorption can be observed by comparing Fig.~\ref{fig:YeIrr}, which shows the electron fraction towards
which the fluid is driven by neutrino absorption, and Fig.~\ref{fig:BarSlice}, which shows the actual electron fraction in the simulations (with both the leakage and the transport scheme). We see that in the disk and right outside of the innermost shocked region, neutrino absorption drives $Y_e$ down (electron antineutrinos strongly dominate electron neutrinos).
The closer to the shocked region a fluid element is, the more neutron-rich it becomes. In the polar regions, on the other
hand, there remain enough electron neutrinos to drive the composition to a higher $Y_e$, at least close to the surface
where the irradiation is the strongest. Finally, outside of the weaker shock at $r\sim 55\,{\rm km}$ (seen on Fig.~\ref{fig:BarSlice}), 
where the electron neutrinos decouple from the disk, the effect of neutrino absorption is also to raise $Y_e$.

\begin{figure}
\flushleft
\includegraphics[width=1.03\columnwidth]{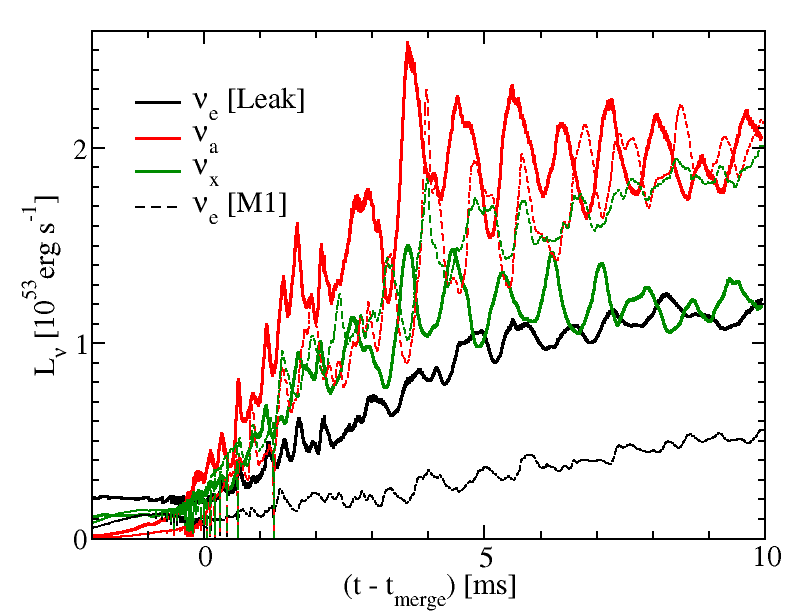}
\caption{Neutrino luminosities for the simulations using the LS220 equation of state and the neutrino leakage scheme
(solid curves), or the neutrino transport scheme (dashed curves). We show the luminosity in electron neutrinos (black),
electron antineutrinos (red), and all heavy-lepton neutrinos and antineutrinos combined (green).}
\label{fig:specieslum}
\end{figure}

Finally, we note that there are significant differences between the leakage and transport simulations even in terms
of the total neutrino luminosity in each species. Figure~\ref{fig:specieslum} shows the total luminosity as a function of time for 
each species for the LS220 simulation, either with neutrino leakage or with neutrino transport. 
We observe differences of as much as
a factor of two. The electron neutrino luminosity, which was already particularly low for the LS220 equation of state when using
the leakage scheme (see 
Table~\ref{tab:nu}),
is a factor of four smaller than the electron antineutrino luminosity when using the transport scheme. 
This can be explained by the fact that a large fraction
of the neutrinos, which the leakage scheme predicts are escaping from the core, are emitted in the
direction of the accretion disk. The disk is optically thick to electron neutrinos, and when using the transport scheme 
those neutrinos are reabsorbed in the disk. Our leakage scheme does not account for this effect, because it allows neutrinos
to move along the path of smallest optical depth (in this case, along the polar axis), instead of following null geodesics when free-streaming. 

We also note that, even though the electron antineutrino luminosity appears to agree well in the leakage and transport simulations, 
this is a mere coincidence. There are indeed two potential sources of disagreement between the leakage and transport results. 
The first is the error in the estimates of the leakage scheme for given physical conditions in the post-merger remnant. 
The second is that the physical properties of the post-merger remnant themselves evolve differently in the leakage and transport
simulations. One important such difference is that the disk is hotter 
in the simulation using neutrino transport, due to neutrino absorption. So the predicted emission rate of $\bar{\nu}_e$ is larger
in the transport simulation. But the leakage scheme neglects neutrino absorption, which causes it to overpredict the total luminosity of
$\bar\nu_e$ (albeit not by as much as for $\nu_e$). In this specific case, the two effects appear to largely cancel each other. But
one can check that they are not negligible by using the leakage scheme to predict the neutrino luminosity at a given time
in the simulation using a transport scheme. The resulting $\bar\nu_e$ luminosity is a factor of two larger than when using the 
transport scheme.

In conclusion, it appears that using a neutrino transport scheme is necessary for reliable predictions regarding the composition
of the outflows, the composition of the disk, or the neutrino luminosity. The latter is particularly important in the post-merger
evolution of BNS mergers because of the complex geometry of neutrino emission, and the impact of neutrino absorption
on the evolution of the temperature and composition of the disk.

\section{Outflows}
\label{sec:outflows}

As discussed in Sec.~\ref{sec:methods}, the simulations performed here use too small a numerical grid to accurately measure
the mass of unbound material leaving the system. For reference, however, we provide in Table~\ref{tab:nu} the main 
properties of the material
which satisfies the approximate unbound condition $h u_t<-1$ as it crosses the outer boundary of our grid, with $u^\mu$ the 4-velocity of the fluid. 
The quantity $h u_t$
is conserved along a fluid line if $\partial_t$ is a Killing vector (see e.g.~\cite{Gourgoulhon2010}), which we assume to be approximately true
far from the remnant and after merger.
In Table~\ref{tab:nu} and the rest of this section, all averages refer to 
density weighted averages, i.e.
\beq
\langle X \rangle = \frac{\int \rho_0 \sqrt{-g} u^t X dV}{\int \rho_0 \sqrt{-g} u^t dV},
\eeq
with $g$ the determinant of $g_{\mu\nu}$ and $dV$ the flat-space volume element.
We note that 
for the total mass ejected the low and high resolution simulations with the LS220 equation of state
give very different results. The difference is due to variations in the small amount of mass which is ejected at the time of merger.
This is the only quantity in our simulations for which the two resolutions disagree, and a clear indicator
that the mass of the dynamical ejecta is unreliable. 

\begin{table}
\caption{
  Properties of the outflows measured within the first $10\,{\rm ms}$ following the merger.
  The three simulations using the LS220 equation of state are the leakage simulation at our
  standard resolution (LS220-L0), the leakage simulation at high resolution (LS220-L1),
  and the simulation using the two-moment transport scheme (LS220-M1).
  $M_{\rm ejecta}$ is the amount of mass which is flagged as unbound ($hu_t<-1$) when
  leaving the computational grid, $\langle S\rangle$ and $\langle Y_e \rangle$ are density-weighted
  averages of the entropy and electron fraction, and $M_{\rm polar}$ is the mass of unbound
  material escaping through the upper and lower boundaries of the computational domain.
}
\label{tab:nu}
\begin{ruledtabular}
\begin{tabular}{|l|r|r|r|r|}
{\rm Name} & $M_{\rm ejecta}$ & $\langle S \rangle$ & $\langle Y_e \rangle$ &  $M_{\rm polar}/M_{\rm ejecta}$\\
\hline
Units & $10^{-4}M_\odot$ & $k_B\,{\rm baryon}^{-1}$ &  &  \\
\hline
SFHo & 5 & 10 & 0.11 & 0.13\\ 
LS220-L0 & 2 & 12 & 0.11 & 0.01\\
LS220-L1 & 13 & 10 & 0.10 & 0.07\\
LS220-M1 & 5 & 21 & 0.20 & 0.56 \\
DD2 & 13 & 11 & 0.11 &  0.20\\
\end{tabular}
\end{ruledtabular}
\end{table}

\begin{figure}
\flushleft
\includegraphics[width=1.03\columnwidth]{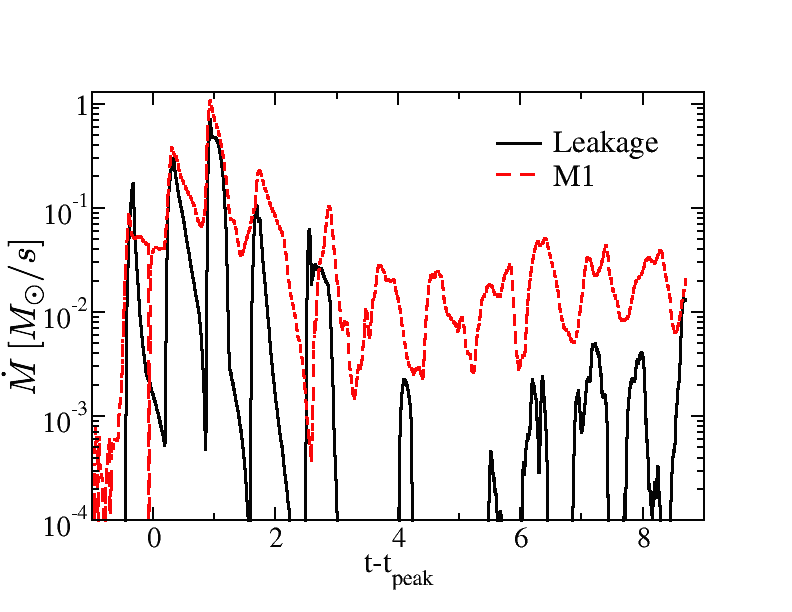}
\caption{Rate at which material flagged as unbound leaves the computational grid in the simulation using the LS220 equation of 
state and the neutrino leakage scheme. We also show the flow of unbound material at the same location in the simulation using
the transport scheme. Note that due to neutrino absorptions at larger distances (the transport simulation uses a larger numerical
grid), the true mass outflows in the simulation using a transport scheme are higher at late times, 
with $\dot{M} \sim 0.04M_\odot\,s^{-1}$. The oscillations in the outflow rates are largely due to the use of a rectangular outer boundary and the asymmetry of the outflows.}
\label{fig:outflow}
\end{figure}

The other notable result is that in all leakage simulations the ejected material has very similar properties:
average specific entropy $\langle s\rangle \sim 10 k_B\,{\rm baryon}^{-1}$, average electron fraction $\langle Y_e \rangle \sim 0.1$,
matter ejection occurring mostly close to the equatorial plane, and no outflows at late times in the simulations.
By comparison, the transport simulation ejects material with a higher electron fraction, twice the average entropy, and shows
the ejection of material in the polar regions at late times in the evolution (see Fig.~\ref{fig:outflow}). This neutrino-driven wind has 
$\dot{M}\sim 0.04M_\odot\,s^{-1}$. If maintained for $50-100\,{\rm ms}$, it could be the dominant source of ejecta in 
the merger (see also~\cite{Dessart2009}). The wind is generally more proton rich
and hotter than the earlier dynamical ejecta, with $\langle Y_e \rangle \sim 0.25$ and 
$\langle s\rangle \sim 35 k_B\,{\rm baryon}^{-1}$. In the M1 simulation, we find $\langle Y_e \rangle \sim 0.18$ 
and $\langle s\rangle \sim 18 k_B\,{\rm baryon}^{-1}$ for the dynamical ejecta (material ejected in the first $\sim 2\,{\rm ms}$ 
following the merger).

The composition and entropy of the ejecta is important for two main reasons. The first is that neutron-rich material
undergoes strong r-process nucleosynthesis, producing mostly heavy elements with number of nucleons $A>120$,
while less neutron-rich material will produce a larger fraction of lower mass elements with $50<A<100$~\cite{Wanajo2014}. The late-time
wind observed here has an electron fraction which is at the limit between the two regimes. Lippuner \& Roberts~\cite{Lippuner2015}
predict that, for the entropy seen in our simulations, the limiting composition is $Y_e\sim 0.23$. This 
suggests the production of a broad range of elements in the neutrino-driven wind, in agreement with what has been observed 
in higher mass neutron
star mergers~\cite{Wanajo2014}, and in the evolution of post-merger accretion disks~\cite{Just2014}. 
The second consequence is that the heavy-elements produced in neutron-rich
material, particularly the lanthanides, have very high photon opacities. Their presence causes the electromagnetic transients powered
by r-process nucleosynthesis to peak on a timescale of a week instead of a few days, and in the infrared instead of the 
optical~\cite{2013ApJ...775...18B}. If there existed a lanthanide free region in the outflows, emission from that region
could power an earlier, optical signal~\cite{Metzger2014}. However, the conditions observed in our simulations do not favor such a scenario.

Finally, the strong wind generated in our simulation with neutrino transport
also causes significant baryon loading of the polar regions, with a measured wind of $\dot{M}\sim 0.04M_\odot\,s^{-1}$
during the last $5\,{\rm ms}$ of the simulation. This could hamper the production of short-gamma ray bursts, 
at least until the collapse of the post-merger neutron star to a black hole, which will not happen for the DD2 equation of state
since the post-merger neutron star is stable in that case.

\section{Conclusions}
\label{sec:concl}

We present a first set of simulations of neutron star mergers using the SpEC code, nuclear-theory based equations
of state, and either a simple neutrino cooling prescription (leakage) or a two-moment grey neutrino transport scheme. 
For this first study,
we focus on equal mass neutron stars with $M_{\rm NS}=1.2M_\odot$, at the low end of the expected range of neutron
star masses. So far, only a few studies with fully general relativistic codes have included the effects of neutrinos with 
leakage~\cite{Neilsen:2014hha,Palenzuela2015} or transport schemes~\cite{Sekiguchi:2015}. All have focused
solely on equal mass systems with $M_{\rm NS}=1.35M_\odot$. These simulations allow us to study the properties
of the post-merger gravitational waveform for BNS mergers with realistic equations of state, the qualitative
properties of the post-merger remnant, the effects of neutrinos on the post-merger evolution, and the impact of the choice
of neutrino treatment on the results of the simulations.

The post-merger gravitational wave signal is known to be dominated by strong peaks at frequencies dependent on the
neutron star equation of state, although there is some disagreement on the interpretation of the peaks and predicted
emission frequency. The strongest post-merger peak, which is associated with the fundamental $l=2$ excitation of the
post-merger neutron star, largely dominates the post-merger signal in our simulations. We find that the location of that 
peak is in good agreement (to better than $\sim100\,{\rm Hz}$) with the predictions of Bauswein et al.~\cite{bauswein:12}, 
which were based on a large
number of simulations using an approximate treatment of gravity. Other general relativistic simulations have recently
found large deviations from those predictions~\cite{Takami:2014zpa}, but for choices of equation of state incompatible
with the expected properties of neutron stars. General relativistic simulations with nuclear-theory based equations
of state of higher mass neutron star mergers~\cite{Palenzuela2015} found agreement at the $10\%$ level
($[200-300]\,{\rm Hz}$ differences) with the fitting formula provided by Bauswein et al.~\cite{bauswein:12}. 
Although the accuracy of that fitting formula remains an open question, our simulations tend to confirm that they perform well for nuclear-theory based equations of state.

We are less optimistic when it comes to the determination of the secondary peaks of the post-merger signal in low-mass
systems. Although we do observe peaks at frequencies close to those recently predicted by 
Bauswein \& Stergoulias~\cite{Bauswein:2015a},
and corroborated by the simulations of Palenzuela et al.~\cite{Palenzuela2015}, we find that their emission is 
limited to a very short time period of $\sim (1-3)\,{\rm ms}$, which causes the peaks to be very broad in frequency space. 
Their detectability when mixed with realistic detector noise may be debatable.

The properties of the post-merger remnant appear to follow the trends observed at higher mass when considering 
low-compactness neutron stars. The post-merger neutron star
is not as strongly heated as in higher mass, more compact systems, but strong $l=2$ modes survive for the entire duration
of the simulation, up to $10\,{\rm ms}$ past merger. When treating the neutrinos through the simple leakage scheme, 
we find good agreement in the basic properties of the remnant (density, temperature, composition) and the observed neutrino luminosity 
with recent results
for higher mass systems with a similar level of physical realism~\cite{Neilsen:2014hha,Palenzuela2015}. 

We also note that the neutron star remnants observed in our simulations are differentially rotating, and that the surrounding disks have rotation profiles close to equilibrium. The post-merger disk is unstable to the
axisymmetric magnetorotational instability, while the neutron star remnant has a complex rotation profile in which parts 
of the neutron star have an angular frequency profile satisfying $d\Omega/dr>0$. These regions 
are stable to the magnetorotational instability, at least in the first $10\,{\rm ms}$ following the merger. 

The inclusion of a more
realistic neutrino transport method significantly modifies the composition of the disk and outflows. Neutrino energy deposition 
also powers
sustained outflows with moderate electron fraction (on average $\langle Y_e \rangle \sim 0.25$) up to the end of the simulations. These outflows are
expected to produce a wide range of elements through r-process nucleosynthesis, as their properties are right at the limit
between neutron-rich material producing mostly heavy elements ($A>120$), and less neutron-rich material producing lower
mass elements. Finally, absorption of electron neutrinos and antineutrinos in the optically thick
disk causes heating in the disk. It also leads to a disagreement in the predicted neutrino luminosity between the leakage scheme, which does not account for absorption, and the transport scheme. None of these differences significantly affect the hydrodynamic 
of the remnant,
but they show the importance of neutrino transport when assessing the mass, composition, and geometry of the outflows, 
the resulting properties of radioactively powered electromagnetic transients, and the result of r-process nucleosynthesis in the outflows.
The fact that neutron star mergers can produce a wide range of elements when neutrino transport is taken into account has been
discussed in more detail by post-processing the results of a higher mass neutron star merger 
simulation~\cite{Wanajo2014,Sekiguchi:2015}, 
and of simulations of post-merger accretion disk models~\cite{Just2014}. Our present simulations confirm these results in low-mass neutron star
mergers. Additionally, they explicitly show the difference between simulations using a simple leakage scheme, and simulations using neutrino transport.

The results presented here are limited by a number of important assumptions in our simulations. The first is the study of equal
mass binaries. Indeed, we know that unequal and equal mass BNS mergers show significant differences. In particular, unequal 
mass mergers eject a larger amount of material~\cite{hotokezaka:13}. The second is the fact that we ignore magnetic fields. Over the short time
scales considered here, magnetohydrodynamics effects are not expected to affect the evolution of neutron star remnant, but could drive additional
outflows from the disk~\cite{Kiuchi2014,Palenzuela2015} . Over longer time scales, magnetic fields would be critical to the spin evolution
of the remnant neutron star, angular momentum transport, heating in the disk, and possibly the formation of relativistic jets and magnetically-driven
outflows. Finally, the relatively small numerical grid on which we evolve the equations of general relativistic hydrodynamics limits our ability to measure
the mass and properties of the outflows accurately. We expect to address these issues in the future. However, we do not expect these assumptions
to significantly affect the main results of this work, i.e.~the properties of the post-merger gravitational wave signal and the importance of the 
neutrino-matter interactions in the first $10\,{\rm ms}$ following the merger.

\acknowledgments
The authors wish to thank Jan Steinhoff for communicating values for the Love number
or the neutron stars used in our simulations, Rodrigo Fernandez and Dan Kasen
for useful discussions over the course of this work, and the members of the SxS collaboration
for their advice and support.
Support for this work was provided
by NASA through Einstein Postdoctoral Fellowship
grants numbered PF4-150122 (F.F.) and PF3-140114 (L.R.) awarded 
by the Chandra X-ray Center, which is operated by the Smithsonian 
Astrophysical Observatory for NASA under contract NAS8-03060; 
and through Hubble Fellowship grant number 51344.001 awarded 
by the Space Telescope Science Institute, which is operated by the Association
 of Universities for Research in Astronomy, Inc., for NASA, under contract NAS 5-26555.
The authors at CITA gratefully acknowledge support from the NSERC
Canada. 
M.D.D. acknowledges support through NSF Grant PHY-1402916.
L.K. acknowledges support from NSF grants PHY-1306125 and AST-1333129 at
Cornell, while the authors at Caltech acknowledge support from NSF
Grants  PHY-1404569, AST-1333520, NSF-1440083, and
NSF CAREER Award PHY-1151197.
Authors at both Cornell and Caltech also thank the Sherman Fairchild Foundation
for their support. 
Computations were performed on the
supercomputer Briar\'ee from the Universit\'e de Montr\'eal, and Guillimin
from McGill University, both managed
by Calcul Qu\'ebec and Compute Canada. The operation
of these supercomputers is funded by the Canada Foundation
for Innovation (CFI), NanoQu\'ebec, RMGA and the Fonds de
recherche du Qu\'ebec - Nature et Technologie (FRQ-NT). Computations were
also performed on the Zwicky cluster at Caltech, supported by the Sherman
Fairchild Foundation and by NSF award PHY-0960291.
This work also used the Extreme Science and Engineering
Discovery Environment (XSEDE) through allocation No. TGPHY990007N,
supported by NSF Grant No. ACI-1053575.

\bibliography{References/References}

\end{document}